\documentclass[prb,showpacs,twocolumn,floatfix]{revtex4}
\usepackage{graphicx}
\begin{document}
\def\rhov{{\mbox{\boldmath{$\rho$}}}}
\def\tauv{{\mbox{\boldmath{$\tau$}}}}
\def\Lambdav{{\mbox{\boldmath{$\Lambda$}}}}
\def\sigmav{{\mbox{\boldmath{$\sigma$}}}}
\def\xiv{{\mbox{\boldmath{$\xi$}}}}
\def\chiv{{\mbox{\boldmath{$\chi$}}}}
\def\rhov{{\mbox{\boldmath{$\rho$}}}}
\def\deltav{{\mbox{\boldmath{$\delta$}}}}
\def\phiv{{\mbox{\boldmath{$\phi$}}}}
\def\piv{{\mbox{\boldmath{$\pi$}}}}
\def\psiv{{\mbox{\boldmath{$\psi$}}}}
\def\oh{{\scriptsize 1 \over \scriptsize 2}}
\def\ot{{\scriptsize 1 \over \scriptsize 3}}
\def\of{{\scriptsize 1 \over \scriptsize 4}}
\def\tf{{\scriptsize 3 \over \scriptsize 4}}
\title{Spin Dynamics of Trimers on a Distorted Kagom\'e Lattice}

\author{A. B. Harris$^1$ and T. Yildirim$^{2,3}$}

\affiliation{ [1]Department of Physics and Astronomy,
University of Pennsylvania, Philadelphia, PA 19104}
\affiliation{ [2]NIST Center for Neutron Research, National Institute of
Science and Technology, Gaithersburg, MD 20899}
\affiliation{ [3]Department of Materials Science and Engineering, University
of Pennsylvania, Philadelphia, PA 19104}
%%% ----------------------------------------------------------------------
\date{\today}

\begin{abstract}
We treat the ground state, elementary excitations, and
neutron scattering cross section for a system of trimers consisting of
three tightly bound spins 1/2 on a distorted Kagom\'e lattice,
subject to isotropic nearest neighbor (usually antiferromagnetic) Heisenberg
interactions.  The interactions between trimers are assumed to be weak
compared to the intra trimer interactions.  We compare the spin-wave
excitation spectrum of trimers with
that obtained from standard spin-wave theory and attribute the
differences at low energy to the fact that the trimer formulation includes exactly
the effects of intra-trimer zero point motion. Application to
existing systems is briefly  discussed.
\end{abstract}
\maketitle

\section{INTRODUCTION}

Frustrated antiferromagnetic systems have received enormous attention
in recent years.\cite{MW,GN,KSB} One limit which has attracted less
attention is that when the frustration is removed by the formation of
strongly coupled three-spin units called 
spin trimers.\cite{JAPS,CB,FURRER1,FURRER2,POD,OKAM,HL}
Early experiments and calculations were performed for
high ($S=5/2)$ spin states of Fe$^{3+}$ and Mn$^{2+}$
by Falk et al.,\cite{FURRER1} and Furrer and G\"udel.\cite{FURRER2}
For $S=1/2$ systems much work has been focused on chain-like systems
consisting of trimers of Cu ions.\cite{JAPS,OKAM,HL} Other
configurations of trimers were studied by Qiu et al.,\cite{CB}
and Podlesnyak et al.\cite{POD} In these works the interactions
between trimers were very weak, so that the energy of the localized
excitations appeared not to depend on wave vector.  In that case,
information on the nature of the excited states of the trimers
was obtained by monitoring the dependence of the
magnitude of the inelastic scattering cross section on wave
vector.  In contrast, here we will consider a system of interacting
spin 1/2 trimers where the excitations have a significant dependence
on wave vector.  We implement perturbation
theory by introducing operators which create or destroy the
the exact excited states of isolated trimers.  In the limit
when the inter-trimer interactions vanish, our calculation
reduces to those of Refs. \onlinecite{FURRER1} and \onlinecite{CB}.

\begin{figure}
\begin{center}
\includegraphics[width=3 in]{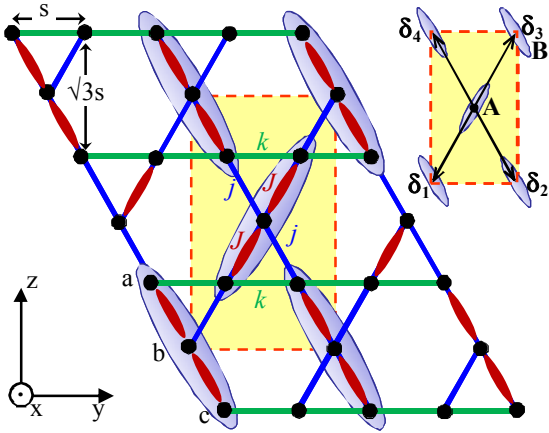}
\caption{\label{KAG} (Color online) 
A distorted Kagom\'e lattice with three isotropic nearest neighbor Heisenberg
interactions; $J$ (red), $j$ (blue)  and $k$ (green). We assume $J$ is 
antiferromagnetic and much larger than $j$ and $k$, yielding spin trimers
(some of which are shown as ellipses)   which consist of three spins connected
by two large interactions, $J$.  The dashed rectangle is the unit cell
containing two trimers, A and B.  The upper right inset shows the four
nearest neighbor vectors of the trimer lattice as given in Eq. (\ref{DELEQ}).
Here $s$ is the nearest neighbor separation between spins on the Kagom\'e
lattice.    The labeling of the three sites within a trimer is $a$, $b$, $c$
in the order of decreasing $z$ coordinate, as shown for a trimer in
the left bottom corner of the unit cell. }
\end{center}
\end{figure}

The system of trimers of spins 1/2 we consider is specified by Fig.
\ref{KAG} where we show the covering of a distorted Kagom\'e lattice
by trimers.  The lattice has the connectivity of a Kagom\'e lattice,
but lacks its three-fold symmetry, so that the nearest neighbor isotropic
exchange interactions assume three values $J$, $j$, and $k$, of
which $J$ is assumed to be dominant. This model may be an appropriate one
for the distorted Kagom\'e system Cu$_2$(OD)$_3$Cl.\cite{SHL1,SHL2}
Even if this system is not an ideal representative of the
model we introduce below, our results may stimulate the search for
better realizations of our model.  The aim of this paper is
to develop a calculation which is correct to leading order in
$j/J$ and $k/J$ and to compare results obtained in this
approximation to standard spin-wave theory, based on the N\'eel state
which treats all the exchange interactions on an equal footing.
We find that there is a one-to-one mapping connecting
the lowest energy manifold of excitations in the two approaches
and that the differences in energies can be understood in terms of
the differing way quantum zero-point motion is treated in
the two approaches.  At higher energy the comparison is more
complicated.  In the trimer approach one does have the higher energy
transverse spin waves of the N\'eel state.  But in addition,
some of the higher energy trimer excitations correspond to bound
states of two or more N\'eel-state spin excitations.  The trimer approach
is clearly superior when the intertrimer interactions
are perturbative, as we assume in this paper.

Briefly, this paper is organized as follows.  In Sec. II 
we give a qualitative overview of the calculation in which
the intertrimer interactions $j$ and $k$ are treated perturbatively
with respect to the strong intratrimer interaction $J$.   
In Sec. III we show
that the low energy manifold of spin waves can be mapped onto
the usual manifold of spin waves, but with an effective
trimer-trimer interaction playing the role of the usual spin-spin
interaction. Here and in  succeeding sections we treat the two
cases when a) the net spins of adjacent trimers are coupled
antiferromagnetically and b) the net spins of adjacent trimers are
coupled ferromagnetically.  In Sec. IV we consider the exciton
spectrum in which trimers are promoted into their nearly localized
excited states. In Sec. V we present results of standard
spin-wave calculations based on the N\'eel state in which
all spins in the ground state have $S_z=1/2$ or $S_z=-1/2$.
In Sec VI we compare the results the spin-wave and perturbative
approaches give for the elastic diffraction pattern.
We attribute the differences
in results to the differences in how quantum zero-point motion
is treated in the two approaches.  In Sec. VII we consider the
inelastic neutron scattering cross section from the entire
spectrum of trimer excitations.  Our results are summarized 
and briefly discussed in Sec. VIII.

\section{OVERVIEW}

\begin{figure}
\begin{center}
\includegraphics[width=2.0 in]{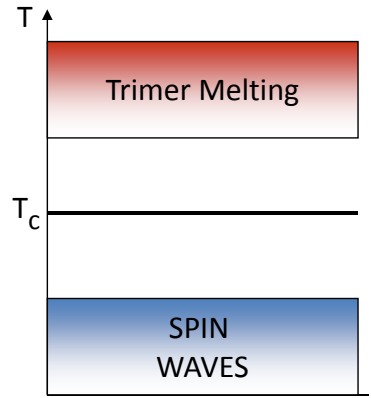} 
\caption{\label{TEMP} (Color online) The phase diagram of the trimer
system as a function of temperature $T$, as discussed in the text.
Long range magnetic order  occurs at $T_c$.  Trimer formation occurs
over the regime for which $T$ is of order $J$.}
\end{center}
\end{figure}

In the magnetically disordered phase of Cu(OD)$_3$Cl (which we
take as the exemplar of our trimer model) the unit cell shown in
Fig. \ref{KAG} contains six Cu spin sites.  In Fig. \ref{TEMP}
we show the phase diagram of the trimer model as a function of
the temperature $T$ when $J$ is much larger than either $j$ or $k$.
When $T$ is large compared to $J$ the spins are essentially uncorrelated.
As $T$ is reduced to become comparable to $J$, one passes through a
regime in which the correlations within spin trimers become well
developed.  In Fig. \ref{TEMP}, this regime is labeled ''trimer melting.''
Below this regime the average spin of the middle site of
the trimer is oppositely oriented to those of the end sites of
the trimer.[\onlinecite{SHL1,SHL2}] However, as long as $T>T_c$, the
spin correlation function between different trimers decays rapidly
as a function of their separation. When $T$ is reduced so as to
be comparable to $j$ and/or $k$, one passes through a phase
transition (at $T=T_c$) below which one has long range spin order.
As we discuss below,  
depending on how $j$ and $k$ compare, the adjacent trimers
can either be organized ferromagnetically or antiferromagnetically.
In either case, the magnetic ordering occurs at zero wave vector.
In other words, the magnetic and paramagnetic unit cells are identical,
each containing two trimer units. As we shall see,
when $T \ll T_c$ the elementary excitations
are identical to spin waves in the usual magnetic systems.

In contrast and as will become apparent, the higher energy trimer
excitons are qualitatively different from the higher energy spin wave
relative to the N\'eel ground state.  To obtain a close correspondence
between the two approaches one should consider trimers consisting of
three large S spins.  In that case, one should pass continuously
between the trimer and N\'eel limits as the ratio of $j$ or $k$
to $J$ is varied.

The Hamiltonian for the system of spins 1/2 which we treat is written as
\begin{eqnarray}
{\cal H} &=&  \sum_{\langle ij \rangle} J_{ij} {\bf S}_i \cdot {\bf S}_j \ ,
\end{eqnarray}
where $\langle ij \rangle$ indicates that the sum is over pairs of
nearest neighbors on the Kagom\'e lattice.  Here we neglect exchange
anisotropy, in particular we do not include the Dzialoshinskii-Moriya
[\onlinecite{DM1,DM2}] interaction, which can be the dominant
anisotropic interaction between spins.[\onlinecite{DM3,DM4}]
The values of the $J$'s
are defined in Fig. \ref{KAG} where the intra-trimer interaction $J$ is
assumed to be dominant.  We will work to leading order in
$j/J$ or $k/J$ which are assumed to be of order $x<< 1$. Thus the expansion
parameter $x$ characterizes the ratio of inter-trimer to
intra-trimer interactions.  When inter-trimer interactions are turned
on, the spectrum of discrete energy levels of isolated trimers
gets broadened into a band of wavelike excitations,
just as happens for atomic energy levels when placed in a solid.  

For this calculation we obviously need the exact eigenfunctions and 
eigenvalues of the trimer Hamiltonian
\begin{eqnarray}
{\cal H}_{\rm T} &=& J {\bf S}_a \cdot {\bf S}_b
+ J {\bf S}_b \cdot {\bf S}_ c\ ,
\end{eqnarray}
where the spins within a trimer are labeled as in Fig. \ref{KAG}.
The total spin $\cal S$ is a good quantum number and assumes the values
3/2 and 1/2. The four states ${\cal S} = 3/2$ are degenerate eigenstates
of ${\cal H}_{\rm T}$ with eigenvalue $J/2$.  The remaining four eigenstates
form two ${\cal S}=1/2$ doublets.  The eigenstates and eigenvalues
of ${\cal H}_{\rm T}$ are listed in Table I.  In the next
section we consider the ground state manifold and in the following
sections we consider excitations to the higher manifolds centered
at energy $J$ and $3J/2$ above the ground state.

Before starting the calculation we should discuss 
when the trimer limit we consider is appropriate.  First of all,
our results show the obvious fact that when the trimers interact
with one another, the single-trimer energy levels get broadened into
a band.  Clearly, a condition for treating isolated trimers as
a starting point, would be that this broadening is small enough
that the bands are separated and qualitatively retain their identity
from the noninteracting limit.  But additionally, in view of the
fact that the trimers will be shown to act as spin 1/2's, one could
question whether this calculation improves the treatment of quantum
zero-point which can be severe for $S=1/2$.  The following qualitative
estimate indicates why the trimer calculation can be useful. Let
us consider excitation relative to the N\'eel state in which spins
are aligned along the $z$-axis.  The perturbation
which creates zero-point motion comes from terms like
$J_{ij}S_-(i) S_+(j)/2$, where the subscript labels the Cartesian component
of spin and the largest such terms are those for which sites $i$ and
$j$ are {\it inside the same} trimer.  This perturbation $V$ connects
the ground state to a state with excitation energy $E=2zJS$, where
$z$, the number of nearest neighbors should be taken to be 1 or 2 because
for each site there are only 1 or 2 strongly coupled neighbors. Thus
$V/E\approx 1/2$.  In contrast, when this type of calculation is
repeated for the trimer state $z$ is now 4, the number of trimer-trimer
nearest neighbors.  Also, perturbative corrections to a system of
isolated trimers are of order $V/E=j/J$, where $j$ is one of the
inter-trimer interactions.  So zero point corrections are less important
for the trimer analog of the N\'eel state than for the usual N\'eel state
in the limit when $j/J$ is small.

\section{Ground-State Excitations}

We first consider the $2^N$-fold degenerate manifold of $N$ trimers
when intertrimer interactions are turned off, so that each trimer
has energy $-J$.  To implement degenerate perturbation theory when
intertrimer interactions are turned on, it is convenient to
map this manifold of states onto the $2^N$ states associated with
a system of $N$ pseudospin 1/2 operators, such that the pseudospin operator
of each trimer is simply the total ground state spin operators ${\cal S}$
of that trimer. For the trimer at position ${\bf R}$ we denote this
pseudospin operator as $\sigmav ({\bf R})$.  Then, any operator within
the ground manifold can be expressed in terms of products of one or
more $\sigmav ({\bf R})$. We then use the wavefunctions in Table I
to express matrix elements of spin operators for individual sites
within the trimer at ${\bf R}$ to $\sigmav({\bf R})$.
For this purpose we label the three spins within a trimer  
as $a$, $b$, and $c$ as in Fig. \ref{KAG}.  Using Table
\ref{EIG} we note that for
an A trimer (for which $\sigmav_z = 1/2$), the expectation value
of the $z$-component of the $k$th spin in the ground state of the trimer
denoted $S_z(k)$ (where $k=a,b,c$) is
\begin{eqnarray}
S_z(a) &=& 1/3 \ , \ \ S_z(b) = - 1/6 \ , \ \ S_z(c) = 1/3 \ .
\end{eqnarray}
This result reflects the fact that the central spin partakes of spin
fluctuations with its two neighbors inside the trimer whereas
an end spin of the trimer has only one neighbor with which to
fluctuate. Below we will discuss the experimental
consequences of this result. In fact, the Wigner-Eckart
theorem\cite{ROSE} indicates that we have, {\it as an operator
equality within the ground manifold}, that
\begin{eqnarray}
{\bf S}(a;{\bf R}) &=& 2 \sigmav ({\bf R})/3 \ , \nonumber \\
{\bf S}(b;{\bf R}) &=& - \sigmav ({\bf R})/3 \ , \nonumber \\
{\bf S}(c;{\bf R}) &=& 2 \sigmav ({\bf R})/3 \ , 
\label{WIGECK} \end{eqnarray}
where ${\bf S}(k;{\bf R})$ is the operator for the $k$th spin in the
trimer whose center is at ${\bf R}$ and, as we have said, the
pseudo-spin operator is identified as the total spin of the trimer:
\begin{eqnarray}
\sigmav ({\bf R}) = \sum_{k=a}^c {\bf S}(k;{\bf R}) \ .
\end{eqnarray}
These equalities make it a trivial matter to write the
inter-trimer interactions in terms of the $\sigmav$'s.
So we see, even without calculation, that the low-energy
spectrum of the trimer system is identical to that of
a system in which each trimer is replaced by an ordinary spin 1/2.
 
\begin{table}
\caption{\label{EIG} Eigenvectors $\psi_n$ and eigenvalues 
$\lambda_n$ of ${\cal H}_{\rm T}$. The states specified by three vertical
arrows give the values of $S_z$ for spins $a$, $b$, and $c$ (reading
from left to right), as shown in Fig. \ref{KAG}.
The index $n$ is only used to label excited states.}
\begin{tabular} { c r c c c}
\hline $n$ & ${\cal S}$ & ${\cal S}_z$ & $\psi_n$ & $\lambda_n$ \\ \hline
\hline $6$&$\frac{3}{2}$ &  $\frac{3}{2}$ & \ \
$| \uparrow , \uparrow , \uparrow \rangle$\ \ & $J/2$ \\
$5$& $\frac{3}{2}$ &  $\frac{1}{2}$
& $[| \uparrow , \uparrow , \downarrow \rangle +
|\uparrow , \downarrow , \uparrow \rangle +
|\downarrow , \uparrow , \uparrow \rangle ] / \sqrt 3$ & $J/2$ \\
$4$& $\frac{3}{2}$ & $-\frac{1}{2}$ &
$[| \uparrow , \downarrow , \downarrow \rangle +
|\downarrow , \uparrow , \downarrow \rangle +
|\downarrow , \downarrow , \uparrow \rangle ] / \sqrt 3$  & $J/2$\\
$3$& $\frac{3}{2}$ & $-\frac{3}{2}$ &
$| \downarrow , \downarrow , \downarrow \rangle $ & $J/2$  \\
$2$& $\frac{1}{2}$ & $\frac{1}{2}$ & 
$[|\uparrow , \uparrow , \downarrow \rangle -
|\downarrow , \uparrow , \uparrow \rangle ] / \sqrt 2$ & \ \ \ $0$\ \ \ \\
$1$& $\frac{1}{2}$ &$-\frac{1}{2}$ &
$[|\uparrow , \downarrow , \downarrow \rangle -
|\downarrow , \downarrow , \uparrow \rangle ] / \sqrt 2$ & $0$ \\
& $\frac{1}{2}$ &$\frac{1}{2}$ &
$[| \uparrow , \uparrow , \downarrow \rangle - 2
|\uparrow , \downarrow , \uparrow \rangle +
|\downarrow , \uparrow , \uparrow \rangle ] / \sqrt 6$ & $-J$ \\
& $\frac{1}{2}$ &$-\frac{1}{2}$ &
$[-| \uparrow , \downarrow , \downarrow \rangle 
+2 |\downarrow , \uparrow , \downarrow \rangle
- |\downarrow , \downarrow , \uparrow \rangle ] / \sqrt 6$ & $-J$ \\
\hline\hline  \end{tabular}
\end{table}

We now consider the ground state and elementary excitations of
the system when weak interactions between trimers are included.
We will assume that all end-to-end exchange interactions between
nearest neighbor trimers assume a common value $k$ and those
between the end of one trimer and the center of its nearest neighbor
assume a common value $j$ as shown in Fig. 1.  This symmetry we have
imposed makes the calculations algebraiclly simple.  If the
intra-trimer and inter-trimer interactions have no special symmetry,
the calculations becomes algebraically more complicated but
are conceptually no more difficult. So here we give results only
for the model of Fig. \ref{KAG}.

\begin{figure}
\begin{center}
\includegraphics[width=3.0 in]{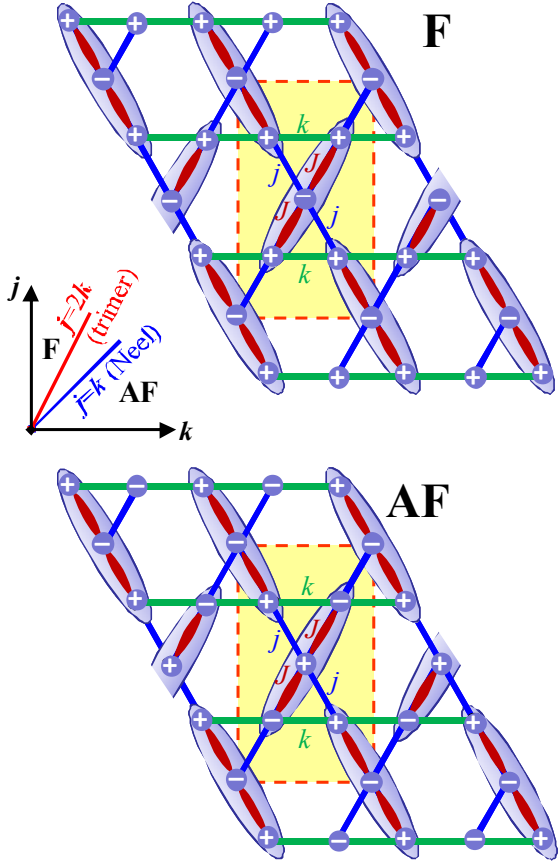}
\caption{\label{PHASE} (Color online) 
As Fig. 1, the ferromagnetic (F) and antiferromagnetic (AF) arrangement
of trimers, with spin orientation indicated by $+$ or $-$.
The inset graph shows the phase diagram of the trimer system in 
the $j$-$k$ plane. The F-AF phase boundary based on the N\'eel state
is at $j=k$ and according to the trimer calculation is at $j=2k$.
The latter calculation is more nearly correct
when $J$ is large compared to $j$ or $k$, whereas the former is more
accurate when $J$ is not large compared to $j$ and $k$.}
\end{center}
\end{figure}

\begin{figure} 
\begin{center}
\includegraphics[width=2.8 in]{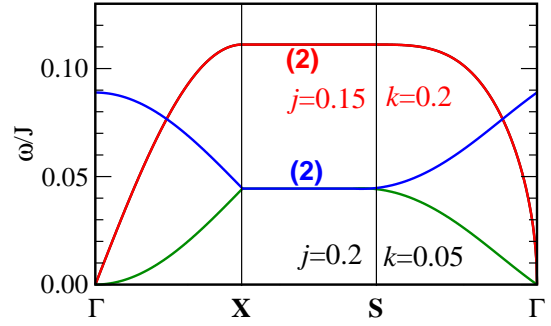} 
\caption{\label{1OM} (Color online) Spectrum of excitation energy,
$\omega ({\bf q})$ within the ground manifold for wave vectors in
special directions.  Here and below we plot the spectra for $J=1$
for wavevectors on the lines joining $\Gamma$ and ${\bf X}$, ${\bf X}$
and ${\bf S}$, and ${\bf S}$ and $\Gamma$, where $\Gamma=(0,0)$,
$X=[\pi/(2s),0)$, and $S=[\pi/(2s),\pi/(2 \sqrt 3 s)$. For
($j=0.15$, $k=0.2$) one has a ferro configuration of trimers and
for ($j=0.2$, $k=0.05$) one has an antiferro configuration of trimers. 
The modes shown here appear only
in the transverse ($+-$) response function.  Modes are nondegenerate
unless labeled ``(2)" to indicate a two-fold degeneracy.}
\end{center}
\end{figure}

We now construct the effective Hamiltonian {\it within the ground
state manifold.} Consider the interaction $V(A,B)$  between
trimers A and B. We use the Wigner-Eckart theorem to express the
spin operators in terms of the pseudo or total spin of the trimer,
as done in Eq. (\ref{WIGECK}). Then one sees that
$V(A,B)$ {\it within the ground manifold} is given by
\begin{eqnarray}
V(A,B)&=& [\sigmav(A) \cdot \sigmav(B)] [ 4k-2j]/9 
\nonumber \\ &\equiv& {\cal J} [\sigmav(A) \cdot \sigmav(B)] \ ,
\label{VEQ} \end{eqnarray}
where
\begin{eqnarray}
{\cal J} &=& (4k-2j)/9 \ .
\label{CALJ} \end{eqnarray}
One sees that the effective exchange interaction between
two nearest neighboring trimers is antiferromagnetic if $2k-j > 0$
and is ferromagnetic if $2k-j <0$.\cite{PB}  Thus the trimer-trimer
interaction can be ferromagnetic even if all the spin-spin interactions
are positive (antiferromagnetic) providing $j>2k$. These configurations
are shown in Fig. \ref{PHASE}.  The elementary excitations within the
ground manifold are those of a rectangular centered lattice.
Then, if the trimers are antiferromagnetically coupled,
standard spin-wave theory\cite{SWAVE} gives the doubly
degenerate spin-wave energy $\omega_\pm ({\bf q})$ as a function
of wave vector ${\bf q}$, for $- \pi/(2s) < q_y < \pi /(2s)$ and
$- \pi /(2 \sqrt 3 s) < q_z < \pi/(2 \sqrt 3 s)$, as
\begin{eqnarray}
\omega({\bf q}) &=& z {\cal J} S \sqrt{ 1 - \gamma({\bf q})^2 } \ , 
\label{EQAF} \end{eqnarray}
where $z=4$ is the number of nearest neighbors, $S=1/2$, and
\begin{eqnarray}
\gamma({\bf q}) &=& (1/z) \sum_\deltav \exp(i {\bf q} \cdot \deltav)
\nonumber \\ &=& \cos (s q_y) \cos (\sqrt 3 s q_z) \ .
\end{eqnarray}
Here $\deltav$ is summed over nearest neighbor vectors between
trimers and $s$ is the nearest neighbor separation in the Kagom\'e
lattice, as in Fig. 1.
If the trimers are ferromagnetically coupled, then
one has two nondegenerate modes whose energy is given by
\begin{eqnarray}
\omega_\pm ({\bf q}) &=& z |{\cal J}| S [ 1 \pm \gamma({\bf q}) ] \ . 
\label{EQF} \end{eqnarray}
Here (in Fig.  \ref{1OM}) and below we give results for $J=1$ for the 
antiferromagnetic configuration of trimers with $j=0.15$ and $k=0.2$
and for the ferromagnetic configuration with $j=0.2$ and $k=0.05$.
Note that transverse ($+-$) modes of the antiferromagnetic
configuration of trimers are doubly degenerate for all wave vectors.
Also here and below note that the spectrum is always two fold
degenerate for wave vectors on the face of the Brillouin zone
[$k_y=\pi /(2s)$] due to the Kramers-like degeneracy from the
two-fold screw axis.[\onlinecite{HEINE}]

%\begin{figure} [ht] 
%\begin{center}
%\includegraphics[width=2.5 in]{NEWSPIN1.eps} 
%\caption{\label{INTFIG} Two nearest-neighboring interacting trimers
%A and B with the labeling of the sites (a,b,c) within the trimer
%together with their spin in terms of their total (or pseudo) spin.}
%\end{center}
%\end{figure}

\section{Exciton Spectrum}

Now we turn to the excitations out of the ground state manifold.

\subsection{Manifold at Energy $J$ for the Antiferro Configuration}

Here we treat the case of antiferromagnetic coupling ( ${\cal J} > 0$).
The situation for this manifold is more complicated than that for
the ground manifold. For the ground manifold we could develop
degenerate perturbation theory for the manifold of $2^N$ states of the
system of $N$ trimers in which each trimer independently occupies
one of its two degenerate ground states.
The result was embodied in an effective Hamiltonian
in which the interactions between nearest neighboring trimers
was given by Eq. (\ref{CALJ}).  For excitations near energy $J$
we might consider the manifold of states in which
one trimer occupies one of the excited states 
of Table I and all the other $N-1$ trimers
are distributed over the two degenerate ground states.
Strictly speaking this involves the solution to a many-body
problem for the states of a spin excitation within the ground
manifold and an exciton at excitation energy $J$ or $3J/2$.
We will not treat this system with this degree of sophistication.
Instead, we will treat the manifold of excited states at 
relative energy $J$ or $3J/2$ when all the background trimers are
confined to their broken symmetry ground state. Thus
our treatment is
limited to the range of temperature $T$ for which $kT \ll {\cal J}$.
We therefore introduce operators $a_n^\dagger({\bf R})$ which take the
trimer at ${\bf R}$ from its ground state to its $n$th excited state,
where the labeling of sites is given in the first column of Table
\ref{EIG}. The Hamiltonian which describes the manifold at energy $J$ is
\begin{eqnarray}
{\cal H}(J) &=& J \sum_{\bf R} \sum_{n=1}^2 n_n({\bf R}) + V(J) \ ,
\label{EQJ} \end{eqnarray}
where ${\bf R}$ is summed over trimer sites and $n_n=a_n^\dagger a_n$.
Within the manifold near energy $J$ the term in Eq. (\ref{EQJ})
proportional to $J$ is a constant and the nature of the band
states is determined solely by the
perturbation $V(J)$, which contains only terms proportional to
$j$ or $k$.  To obtain results to leading order in the expansion
parameter $x$, the perturbation $V(J)$ is thus restricted to terms which
conserve the unperturbed energy $J$.  Accordingly, the most general such
form of $V(J)$ is
\begin{eqnarray}
V(J) &=& \sum_{{\bf R}, {\bf R}'} \sum_{n,m=1}^2
c_{nm}({\bf R},{\bf R}') a_n^\dagger({\bf R}) a_m({\bf R}') + ... 
\label{EQVJ} \end{eqnarray}
where the dots denote terms containing $p$ creation operators (all with
indices in the range 1,2) and $p$ analogous destruction operators.
Since we only consider nearest-neighbor interactions, we set
\begin{eqnarray}
{\bf R}' = {\bf R}_n ={\bf R} + \deltav_n \ ,
\label{RNEQ}
\end{eqnarray}
where $\deltav_n$
are the nearest neighbor intertrimer displacements shown in Fig. \ref{KAG}:
\begin{eqnarray}
\deltav_1 &=& -s \hat j - \sqrt 3 s \hat k \ , \hspace{0.3 in}
\deltav_2 = s \hat j - \sqrt 3 s \hat k \nonumber \\
\deltav_3 &=& s \hat j + \sqrt 3 s \hat k , \hspace{0.3 in} 
\deltav_4 = -s \hat j + \sqrt 3 s \hat k \ .
\label{DELEQ} \end{eqnarray}

The effect of these $2p$th order terms in Eq. (\ref{EQVJ}) on the mode
energies is proportional to the $(p-1)$th power of the density of 
excitations.  Since we assume that $kT \ll J$, this density is small 
and we keep only the terms with $p=1$.  In addition  we 
ignore the kinematic constraint which allows one to map the finite number
of trimer states onto the infinite number of bosonic 
states.[\onlinecite{FJD}] The discussion
for the band at energy $3J/2$ is completely analogous to that for
energy $J$ and the analogous result holds for
that case. So the band states are completely determined by the
matrix $c_{n,m}({\bf R},{\bf R}')$, or, as will shall see, by
its Fourier transform which is a 4$\times$4 matrix for the band
at energy $J$ and an 8$\times$8 matrix for the band at energy $3J/2$.
To explicitly determine $V(J)$ we must express the spin Hamiltonian
in terms of the creation and annihilation operators of Eq. (\ref{EQVJ}).
The spin interaction between the $k$th spin of an up trimer at ${\bf R}$
and the $k'$th spin of a down trimer at ${\bf R}'$ is
\begin{eqnarray}
&& {\bf S} (k;{\bf R}) \cdot {\bf S} (k';{\bf R}') = S_z (k;{\bf R})
S_z (k';{\bf R}') \nonumber \\ && \ \  + [S_+(k;{\bf R}) S_-(k';{\bf R}')
+ S_-(k;{\bf R}) S_+(k';{\bf R}')]/2 \ ,
\label{SDOTSEQ} \end{eqnarray}
Since $S_+$ and $S_-$ each involve at least one creation or
annihilation operator, to construct the boson Hamiltonian,
we need only keep terms in these operators
which are linear in the creation or destruction operators. In
contrast, since $S_z$ has a nonzero value in the ground state,
we also need to keep terms in $S_z$ which involve one creation operator
and one destruction operator within the band. These considerations
will be used implicitly below to limit the complexity of the mapping
from spins to bosons.

For the case of an ``up" trimer at ${\bf R}$ (one whose ground state has
${\cal S}_z=1/2$ and which we refer to as an ``A" trimer) we find 
(keeping only terms linear in the boson operators) that
\begin{eqnarray}
S_- (a;{\bf R}) &=& a_1^\dagger({\bf R})/\sqrt 3
- a_4^\dagger({\bf R}) / \sqrt {18} \nonumber \\ &&
+ a_6({\bf R})/ \sqrt 6 \nonumber \\
S_- (b;{\bf R}) &=& 2a_4^\dagger({\bf R})/ \sqrt {18}
- 2a_6({\bf R})/ \sqrt 6 \nonumber \\
S_- (c;{\bf R}) &=& -a_1^\dagger({\bf R})/\sqrt 3
- a_4^\dagger({\bf R})/ \sqrt {18} \nonumber \\ &&
\ \ + a_6({\bf R})/ \sqrt 6 \ .
\label{EQAA} \end{eqnarray}
The expression for $S_+(k,{\bf R})$ are obtained by Hermitian conjugation.  
To determine the bosonic equivalent of $S_z$ we write
\begin{eqnarray}
S_z &=a_0 + \sum_{nm} a_{nm} a_n^\dagger a_m \ .
\label{SZEQ} \end{eqnarray}
To determine the coefficients we require that the two representations
lead to the same matrix elements.  Thus if 0 labels the ground state
(i. e. whichever of the $-J$ states of the trimer is the ground state),
then, by taking matrix elements of both sides of Eq. (\ref{SZEQ})
we get
\begin{eqnarray}
a_0 &=& \langle 0 | S_z | 0 \rangle \ , \ \ \ \
a_{n,m}  = \langle n | S_z | m \rangle \ , \ \ \ n \not= m \nonumber \\
a_0 &+& a_{n,n} =  \langle n | S_z | n \rangle \ , \ \ \ n \not= 0 \ .
\label{SZEQ2} \end{eqnarray}
So for diagonal elements we must remember to subtract off the ground
state value when identifying the bosonic matrix elements $a_{nn}$.  Thus
\begin{eqnarray}
&& S_z(a;{\bf R})= 1/3 + a_2^\dagger({\bf R})/\sqrt{12}
- a_5^\dagger({\bf R})/ \sqrt {18} \nonumber \\ && + a_2({\bf R})/\sqrt{12}
- a_5({\bf R})/ \sqrt {18} -[ 2n_1({\bf R}) + 2n_2({\bf R})
\nonumber \\ &&  + 5 n_3({\bf R})
+ 3n_4({\bf R}) + n_5({\bf R}) - n_6({\bf R})]/6 \ , \nonumber \\
&& S_z (b;{\bf R}) = -1/6 + 2a_5^\dagger({\bf R})/\sqrt {18}
\nonumber \\ && + 2a_5({\bf R})/\sqrt {18} - n_1({\bf R})/3 
+ 2 n_2({\bf R}) - n_3({\bf R}) /3 \nonumber \\ &&
+ n_5({\bf R})/3 + 2 n_6({\bf R})/3 \ , \nonumber \\
&& S_z (c;{\bf R}) = 1/3 - a_2^\dagger({\bf R})/\sqrt {12} 
- a_5^\dagger({\bf R})/ \sqrt {18} \nonumber \\ &&
- a_2({\bf R})/\sqrt {12} - a_5({\bf R}) / \sqrt {18} 
-[ 2n_1({\bf R}) + 2 n_2({\bf R}) \nonumber \\ && +5 n_3({\bf R})
+3n_4({\bf R}) + n_5({\bf R}) - n_6({\bf R}) \ .
\label{EQAB} \end{eqnarray}
Here we needed to keep $a_p^+a_p\equiv n_p$ terms in view of 
Eqs. (\ref{SDOTSEQ}) and (\ref{SZEQ2}).

For the case of a ``down" trimer (one whose ground state has ${\cal S}_z=-1/2$
at ${\bf R}$ and which we refer to as a ``B" trimer) we similarly find that
\begin{eqnarray}
&& S_+(a;{\bf R}) = a_2^\dagger({\bf R})/\sqrt 3
+ a_5^\dagger({\bf R})/ \sqrt {18}
\nonumber \\ && - a_3({\bf R})/ \sqrt 6 \nonumber \\
&& S_+(b;{\bf R}) = - 2a_5^\dagger({\bf R})/\sqrt{18}+2a_3({\bf R})
/\sqrt 6 \nonumber \\
&& S_+(c;{\bf R}) = -a_2^\dagger({\bf R})/\sqrt 3 
+ a_5^\dagger({\bf R})/\sqrt {18}
\nonumber \\ && \ \  - a_3({\bf R})/ \sqrt 6 \ , \nonumber \\
&& S_z(a;{\bf R})= -1/3 - a_1^\dagger({\bf R})/\sqrt{12}
\nonumber \\ && -a_4^\dagger({\bf R})/\sqrt{18}
- a_1({\bf R})/\sqrt {12} - a_4({\bf R}) / \sqrt {18} \nonumber \\ && 
+ n_1({\bf R})/3 + n_2({\bf R})/3 - n_3({\bf R})/6
\nonumber \\ && + n_4({\bf R})/6 + n_5({\bf R})/2 
+ 5 n_6({\bf R}) /6 \ , \nonumber \\
&& S_z (b;{\bf R}) = 1/6 + 2a_4^\dagger({\bf R})/\sqrt {18}
\nonumber \\ && + 2a_4({\bf R})/\sqrt {18}
- 2 n_1({\bf R})/3 + n_2({\bf R})/3 \nonumber \\ &&
- 2n_3({\bf R})/3 - n_4({\bf R})/3 + n_6({\bf R})/3 \ , \nonumber \\
&& S_z (c;{\bf R}) = -1/3 + a_1^\dagger({\bf R})/\sqrt {12}
\nonumber \\ && -a_4^\dagger({\bf R})/\sqrt {18} 
+ a_1({\bf R})/\sqrt {12} - a_4({\bf R}) / \sqrt {18} \nonumber \\ &&
+ n_1({\bf R})/3 + n_2({\bf R})/3 - n_3({\bf R})/6
\nonumber \\ && + n_4({\bf R})/6 + n_5({\bf R})/2
+ 5 n_6({\bf R})/6 \  .
\end{eqnarray}

The next step is to write the interaction between trimers in terms of
boson operators.  Since we treat here the case when the trimers are
antiferromagnetically coupled, all interactions couple an up (A)
trimer to a down (B) trimer.  Since we treat only nearest
neighbor interactions, we need consider only interactions between
an up trimer at ${\bf R}$ and one of its four down neighbors at
${\bf R}\pm \deltav_1$ and ${\bf R} \pm \deltav_2$. For the excitations
band near energy $J$ the boson Hamiltonian is obtained in Appendix A.
We define the Fourier transformed variables as
\begin{eqnarray}
a_{n,A}^\dagger ({\bf k})= N^{-1/2} \sum_{{\bf R} \in {\rm A}} e^{i {\bf k} 
\cdot {\bf R} } a_n^\dagger ({\bf R}) \nonumber \\
a_{n,B}^\dagger ({\bf k})= N^{-1/2} \sum_{{\bf R} \in {\rm B}} e^{i {\bf k} 
\cdot {\bf R}} a_n^\dagger ({\bf R} ) \ ,
\end{eqnarray}
where $N$ is the total number of unit cells in the system.
The quadratic Hamiltonian is of the canonical form: 
${\cal H} = \sum_{\bf q} {\cal H}_{\bf q}$,
where ${\bf q}$ is the wave vector and because we need consider only terms
which conserve the unperturbed energy $J$,
\begin{eqnarray}
{\cal H}_{\bf q} &=& \sum_{n,n';\tau,\tau'} 
A_{st}({\bf q}) a_s^\dagger({\bf q})
a_t({\bf q}) \ ,
\end{eqnarray}
where $s \equiv (n,\tau)$ and $t \equiv (n',\tau')$.  

According to Table \ref{EIG}, excitations near energy $J$ involve states
1 and 2 of the two spins in the unit cell, whereas excitations near energy
$3J/2$ involve states 3, 4, 5, snf 6 of the two spins in the unit cell.
For excitations near energy $J$ we write
\begin{eqnarray}
{\bf A} &=& J  {\cal I} + k {\bf A}_k + j {\bf A}_j \ ,
\label{AEQ} \end{eqnarray}
where ${\cal I}$ is the $4 \times 4$ unit matrix and
Eq. (\ref{EQA7}) of Appendix A implies that
\begin{eqnarray}
{\bf A}_k  &=& \frac{1}{9} \left[ \begin{array} {c c c c}
4 & 0 & 0 & 0 \\
0 & 4 & 3 \gamma ({\bf q}) & 0 \\
0 & 3 \gamma ({\bf q}) & 4 & 0 \\
0 & 0 & 0 & 4 \\ \end{array} \right] \ ,
\end{eqnarray}
and
\begin{eqnarray}
{\bf A}_j  &=& \frac{1}{9} \left[ \begin{array} {c c c c}
1 & 0 & 0 & 0 \\
0 & - 5 & 0 & 0 \\
0 & 0 & -5 & 0 \\
0 & 0 & 0 & 1 \\
\end{array} \right] \ .
\end{eqnarray}
The rows and columns of the matrices ${\bf A}$ are labeled in
the order $(1,A)$, $(2,A)$,  $(1,B)$, $(2,B)$.

Thus the creation operators for the normal modes are
$a_{1,A}({\bf q})^\dagger$, $a_{2,B}({\bf q})^\dagger$, and
\begin{eqnarray}
\rho_\pm^\dagger &=& [a_{1,B}({\bf q})^\dagger \pm 
a_{2,A} ({\bf q})^\dagger]/\sqrt 2 \ ,
\label{RHOEQ} \end{eqnarray}

\begin{figure} [ht] 
\begin{center}
\includegraphics[width=3.2 in]{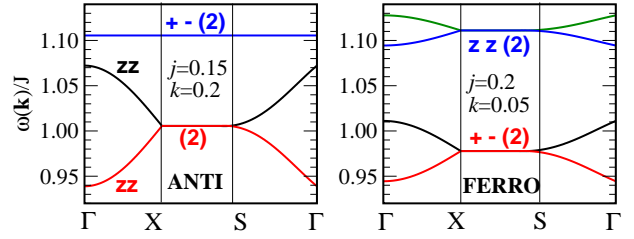} 
\caption{\label{2OM} (Color online) As Fig. \ref{1OM}, but for excitations
in the manifold near energy $J$ for the antiferro (left) and ferro (right) 
configurations. The curve labeled "$+-$"
indicates the energy in the transverse ($+-$) response function and
those labeled "zz" are the energies in the longitudinal ($zz$) response
function. The numbers in parentheses indicate the degeneracy of the mode.}
\end{center}
\end{figure}
with associated eigenenergies
\begin{eqnarray}
\omega_{1A} ({\bf q}) &=& \omega_{2B}({\bf q}) = J + (4k+j)/9
\label{1A2B} \end{eqnarray}
and
\begin{eqnarray}
\omega_\pm ({\bf q}) &=& J + (4k-5j)/9 \pm (k/3) \gamma ({\bf q}) \ . 
\label{PMEQ} \end{eqnarray}
These results are shown in Fig. \ref{2OM}. For an A (up) trimer
$a_{1,A}^\dagger$ corresponds to $S_-$ and $a_{2,B}^\dagger$
corresponds to $S_+$ for a B (down) trimer. So these operators
create transverse excitations and similarly one sees that
$\rho_{\pm}^\dagger$ create a longitudinal excitation.
It may be surprising that, unlike for a N\'eel antiferromagnet,
the longitudinal excitations exhibit dispersion, but the transverse
ones do not.  However, note that for a N\'eel antiferromagnet
the dispersion comes from $a^\dagger a^\dagger$ terms which here
are eliminated since they do not conserve the large unerperturbed
energy.

\subsection{Manifold at Energy $J$ for the Ferro Configuration}

The calculations for the ferro configuration (in which all trimers
start in their `up' ground state) are similar and
are done in Appendix C.  In terms of Fourier transformed variables
Eq. (\ref{EQC7}) implies, in the notation of Eq. (\ref{AEQ}), that
\begin{eqnarray}
{\bf A}_j  &=& \frac{1}{9} \left[ \begin{array} {c c c c}
-1 & 0 & 0 & 0 \\
0 & 5 & 0 & 0 \\
0 & 0 & -1 & 0 \\
0 & 0 & 0 & 5 \\ \end{array} \right] \ ,
\end{eqnarray}
\begin{eqnarray}
{\bf A}_k  &=& \frac{1}{9} \left[ \begin{array} {c c c c}
-4 & 0 & -6 \gamma ({\bf q}) & 0 \\
0 & -4 & 0 & - 3 \gamma ({\bf q}) \\
-6 \gamma({\bf q}) & 0 & -4  & 0 \\
0 & -3 \gamma({\bf q}) & 0 & -4 \\ \end{array} \right] \ ,
\nonumber \\
\end{eqnarray}
where the rows and columns are labeled in the order $(A,1)$, $(A,2)$,
$(B,1)$, $(B,2)$.  The eigenvalues give the mode energies:
\begin{eqnarray}
\omega_{1,2} &=& J -j/9 -4k/9 \pm 2k\gamma({\bf q})/3 \nonumber \\
\omega_{3,4} &=& J + 5j/9 -4k/9 \pm k \gamma({\bf q})/3 \ .
\end{eqnarray}

These mode energies are shown for high symmetry wave vectors
in Fig. \ref{2OM}. Since $a_{A1}^\dagger$ or $a_{B1}^\dagger$
connects the up ground state to a state with ${\cal S}_z=-1/2$,
these operators correspond to an $S_-$. Thus we identify
$\omega_{1,2}$ as energies of transverse excitations and
$\omega_{3,4}$ as energies of longitudinal excitations as 
indicated in Fig. \ref{2OM}.

\subsection{Manifold at Energy $3J$/2 for the Antiferro Configuration}

\begin{figure}
\begin{center}
\includegraphics[width=3 in]{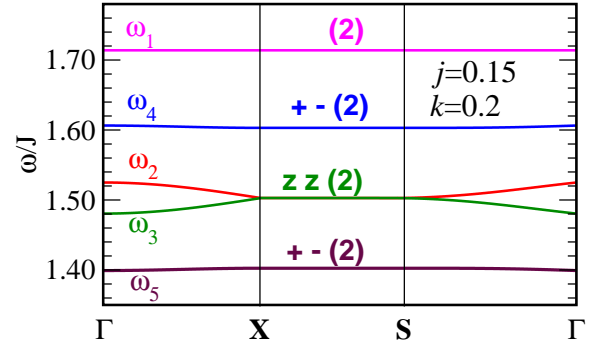} 
\caption{\label{3OM} (Color online) As Fig. \ref{2OM}, but for modes near
energy $3J/2$ for the antiferro configuration.  The highest-energy
mode is not accessible in linear (in ${\bf S}$) response theory.}
\end{center}
\end{figure}

Here we adopt the same simplified approximation in which
trimers not in excited states remain in their N\'eel state.
Then, to leading order in the inter-trimer interactions,
we only keep terms which are quadratic in the variables 3, 
4, 5, and 5 and which conserve the total number of excitations.
Thus analagously to Eq. (\ref{EQJ}) we write
\begin{eqnarray}
{\cal H}(3J/2) &=& (3J/2)\sum_{\bf R} \sum_{n=3}^6
n_n({\bf R}) + V(3J/2) \ ,
\end{eqnarray}
The evaluation of $V(3J/2)$ for the antiferro configuration is
given in Eq. (\ref{EQB7}) of Appendix B.  In the notation of Eq. (\ref{AEQ}),
where we label the rows and columns of the matrices in the order
3A, 6B, 4A, 3B, 5A, 4B, 6A, 5B, that result implies that
\begin{eqnarray}
{\bf A}_k  &=& \frac{1}{9} \left[ \begin{array} {c c c c c c c c}
10& 0 & 0 & 0 & 0 & 0 & 0 & 0 \\
0 &10 & 0 & 0 & 0 & 0 & 0 & 0 \\
0 & 0 & 6 & X & 0 & 0 & 0 & 0 \\
0 & 0 & X &-2 & 0 & 0 & 0 & 0 \\ 
0 & 0 & 0 & 0 & 2 & Y & 0 & 0 \\
0 & 0 & 0 & 0 & Y & 2 & 0 & 0 \\
0 & 0 & 0 & 0 & 0 & 0 &-2 & X \\
0 & 0 & 0 & 0 & 0 & 0 & X & 6 \\\end{array} \right] \ ,
\label{3FA} \end{eqnarray}
\begin{eqnarray}
{\bf A}_j  &=& \frac{1}{18} \left[ \begin{array} {c c c c c c c c}
-1 & 0 & 0 & 0 & 0 & 0 & 0 & 0 \\
0 &-1 & 0 & 0 & 0 & 0 & 0 & 0 \\
0 & 0 & -3& T & 0 & 0 & 0 & 0 \\
0 & 0 & T &-7 & 0 & 0 & 0 & 0 \\ 
0 & 0 & 0 & 0 &-5 & U & 0 & 0 \\
0 & 0 & 0 & 0 & U &-5 & 0 & 0 \\
0 & 0 & 0 & 0 & 0 & 0 &-7 & T \\
0 & 0 & 0 & 0 & 0 & 0 & T &-3 \\\end{array} \right] \ ,
\label{3FB} \end{eqnarray}
where
\begin{eqnarray}
T&=& - 4 \sqrt 3 \gamma({\bf q})  \ , \ \ \ U=-8 \gamma({\bf q})
\ , \nonumber \\ X &=& \sqrt 3 \gamma({\bf q}) \ ,
\ \ \ Y= 2\gamma({\bf q}) \ .
\end{eqnarray}
Thus we have the mode energies, with their
degeneracies in parentheses:
\begin{eqnarray}
\omega_1 &=& 3J/2 + 10 k/9 - j/18 \ (2) \ , \nonumber \\
\omega_{2,3} &=& 3J/2+ 2k/9 -5j/18 \pm (4j-2k)\gamma({\bf q}) / 9 \ \ (1)
\nonumber \\
\omega_{4,5} &=& 3J/2  + 2k/9 -5j/18 \nonumber \\ 
& \pm & \sqrt{ (4k + j)^2 +3 (k-2j)^2\gamma({\bf q})^2}/9 \ \ (2) \ .
\label{MODESJA} \end{eqnarray}
We determine the polarization of the modes as follows. The mode
$\omega_1$ involves state 3A which has ${\cal S}_{A,z}=-3/2$ or
state 6B which has ${\cal S}_{B,z}=3/2$ and is therefore not
accessible via a single spin operator from the 
${\cal S}_{A,z}=1/2$, ${\cal S}_{B,z}=-1/2$ ground state. The modes
$\omega_2$ and $\omega_3$ arise from states 5A and 4B. State 5A has
${\cal S}_{A,z}=1/2$, which is activated from the ${\cal S}_{A,z}=1/2$
ground state by an ${\cal S}_{A,z}$ operator and state 4B has 
${\cal S}_{B,z}=-1/2$ which is activated from the ${\cal S}_{B,z}=-1/2$
ground state by an ${\cal S}_{B,z}$ operator.  The modes $\omega_4$ and
$\omega_5$ arise from states 4A, 3B, 6A, or 5B. State
4A has ${\cal S}_{A,z}=-1/2$, which is activated by an ${\cal S}_{A,-}$
operator and state 5B has ${\cal S}_{B,z}=1/2$ which is activated by
an ${\cal S}_{B,+}$ operator. States 3B or 6A lead to similar results.
These modes (with their polarizations) are shown in Fig. \ref{3OM}.

\subsection{Manifold at energy $3J$/2 for Ferro Configuration}

\begin{figure}
\begin{center}
\includegraphics[width=3 in]{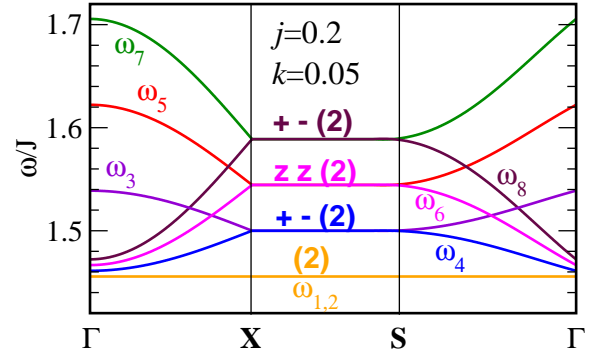} 
\caption{\label{A3OM} (Color online) As Fig. \ref{2OM}, but for excitations
in the manifold near energy $1.5J$ for the ferro configuration from
Eq. (\ref{EQFERRO}).  The lowest-energy mode is not accessible in
linear (in ${\bf S}$) response theory.}
\end{center}
\end{figure}

The result of the calculation of $V(3J/2)$ for the ferro configuration
is given in Eq. (\ref{EQD7}) of Appendix D, which implies,
in the notation of Eq. (\ref{AEQ}), that
\begin{eqnarray}
{\bf A}_k  &=& \frac{1}{9} \left[ \begin{array} {c c c c c c c c}
-10& 0& 0 & 0 & 0 & 0 & 0 & 0 \\
0 &-6 & 0 & 0 & 0 & 1 & 0 & 0 \\
0 & 0 &-2 & 0 & 0 & 0 & 2 & 0 \\
0 & 0 & 0 & 2 & 0 & 0 & 0 & 3 \\ 
0 & 0 & 0 & 0 &-10& 0 & 0 & 0 \\
0 & 1 & 0 & 0 & 0 &-6 & 0 & 0 \\
0 & 0 & 2 & 0 & 0 & 0 &-2 & 0 \\
0 & 0 & 0 & 3 & 0 & 0 & 0 & 2 \\\end{array} \right] \ ,
\label{3AA} \end{eqnarray}
\begin{eqnarray}
{\bf A}_j  &=& \frac{1}{18} \left[ \begin{array} {c c c c c c c c}
1 & 0 & 0 & 0 & 0 & 0 & 0 & 0 \\
0 & 3 & 0 & 0 & 0 & 4 & 0 & 0 \\
0 & 0 & 5 & 0 & 0 & 0 & 8 & 0 \\
0 & 0 & 0 & 7 & 0 & 0 & 0 &12 \\ 
0 & 0 & 0 & 0 & 1 & 0 & 0 & 0 \\
0 & 4 & 0 & 0 & 0 & 3 & 0 & 0 \\
0 & 0 & 8 & 0 & 0 & 0 & 5 & 0 \\
0 & 0 & 0 &12 & 0 & 0 & 0 & 7 \\\end{array} \right] \ ,
\label{3AB} \end{eqnarray}
where the rows and column are labeled in the order 3A, 4A, 5A, 6A,
3B, 4B, 5B, 6B.  We thereby find the mode energies to be
\begin{eqnarray}
\omega_{1,2} &=& 1.5J + \frac{j-20k}{18} \ , \nonumber \\
\omega_{3,4} &=& 1.5J + \frac{3j-12k}{18} \pm \frac{k-2j}{9}\gamma({\bf q})
\ , \nonumber \\
\omega_{5,6} &=& 1.5J + \frac{5j-4k}{18} \pm \frac{2k-4j}{9}\gamma({\bf q})
 \ , \nonumber \\
\omega_{7,8} &=& 1.5J + \frac{7j+4k}{18} \pm \frac{k-2j}{3}\gamma({\bf q}) \ .
\label{EQFERRO} \end{eqnarray}
The determination of the polarization of the mode is done as we did
for the modes of Eq. (\ref{MODESJA}).  The results are shown in
Fig. \ref{A3OM}.

\section{N\'EEL SPIN WAVES}

\begin{figure}
\begin{center}
\includegraphics[width=3.1 in]{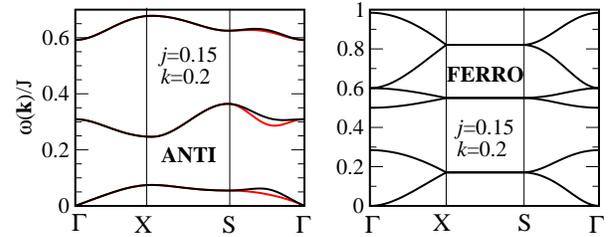} 
\caption{\label{NEEL1} (Color online) As Fig. \ref{1OM}. N\'eel (transverse)
spin-wave spectrum for the anti (left) and ferro (right) configurations.  
Note that the two-fold degeneracy of the antiferromagnetic spectrum is broken
along the low symmetry ${\bf S}$-$\Gamma$ line.}
\end{center}
\end{figure}

In this section we compare the results obtained above with
those from ordinary spin-wave theory. In Fig.  \ref{NEEL1} 
we show the 6 branches of 
transverse excitations from the N\'eel ground state.

Note that apart from the lowest manifold, the two approaches
lead to quite different spectra.  As we showed above, the lowest
manifold of trimer excitations is obtained by an exact mapping
onto a N\'eel spin spectrum.  One sees that for the anti configuration
the energy scale of
the lowest branch of spin waves is significantly larger for the
trimer approach than for the N\'eel approach.  This is because
the trimer approach takes better account of quantum zero point
motion that does the N\'eel approach.  It is well known that
zero point fluctuations tend to increase the spin-wave energies.
This is shown by exact calaculations for one dimensional spin 
chains[\onlinecite {CHAINS}] and by perturbative calculations
for three dimensional systems.[\onlinecite{PERT}] In contrast,
for the ferro configuration the opposite effect occurs
because the energies are proportional to the spin magnitudes.

\section{NEUTRON DIFFRACTION}

\begin{figure}[h*]
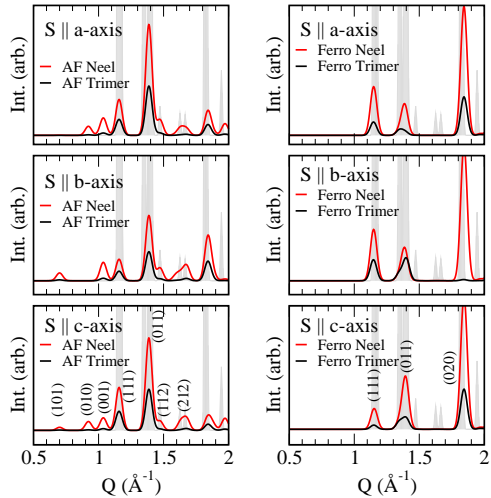

\begin{center}
\includegraphics[width=3cm]{AF_int.eps} $\; \;$
\includegraphics[width=3cm]{Ferro_int.eps} 
\caption{\label{FIG8} Elastic magnetic Bragg peaks for different
spin alignments for trimer and N\'eel models for the AF (left) and
Ferro (right) configurations, respectively. 
The gray lines in the background show the nuclear scattering.  }
\end{center}
\end{figure}

Some aspects of neutron diffraction have been discussed by
Furrer et al.\cite{FURRER1} and by Qiu et al.\cite{CB} Here
we discuss briefly the difference between the diffraction spectrum
of the trimer system and that of the associated N\'eel state.
The elastic magnetic scattering intensity is proportional to 
\begin{equation}
\frac{d \sigma}{d \Omega} \approx \sum_{{\bf G}}
( |{\bf F}({\bf Q}) |^2 - | \hat{\bf Q} \cdot {\bf F}({\bf Q})|^2 ) 
\delta({\bf Q} - {\bf G}),
\end{equation}
where ${\bf G}$ is summed over all reciprocal lattice vectors and the
magnetic vector structure factor ${\bf F}$ is
\begin{equation}
{\bf F}({\bf Q}) \approx \sum_{\tau} \langle {\bf S}_{\tau} \rangle
e^{i {\bf Q} \cdot {\bf \tau}},
\end{equation}
where ${\bf \tau}$ are the copper spin-positions given in Table~\ref{CRYST}
and $\langle {\bf S}_{\tau} \rangle$ is the thermal average of the spin
at site ${\bf \tau}$.  For the N\'eel model, we take the spin-values as 0.5 while
for the trimer model it is 1/6 and 1/3 as shown in Table~\ref{CRYST}.
To simplify the presentation we do not discuss the atomic form factor and the
Debye-Waller factor. The magnetic elastic diffraction intensities
(apart from the thermal and magnetic form factors) are summarized 
in Figure~\ref{FIG8} for different collinear spin configurations
along the crystal axes for both the trimer and N\'eel models, 
including AF and Ferro spin configurations. As expected, there are
significant differences between the antiferro and ferro ordered trimer 
configurations. Also, for a given spin-configuration, the
trimer model is significantly different than the N\'eel model.
Due to smaller spin values in the trimer phase, the intenties
are much weaker. Hence, observation of the magnetic Bragg peaks
would be much more difficult in the trimer phase than for the N\'eel model.
Other than this difference, there are other differences at various
scattering angle and it may be possible to distinguish the
N\'eel and trimer model experimentally. In Figure ~\ref{FIG8}, we also
show nuclear scattering, which has some overlap with the strongest
magnetic peaks. The unique magnetic peaks are at low angle and
there are only a few of them.

\begin{table} [h*]
\caption{ Structure parameters[\onlinecite{WILLS}] \label{CRYST}
(very similar results are given in Ref. \onlinecite{ACTA}) for the
distorted Kagom\' e system Cu$_2$(OD)$_3$Cl\cite{SHL1,SHL2}.  
Here $x$, $y$, and $z$ are the P2$_1$c fractional coordinates with respect
to axes  ${\bf a}= 9.1056\AA$, ${\bf b} =6.8151\AA$, and 
${\bf c} = 11.829\AA$, with $\beta=30.825^{\rm o}$. We choose the
P2$_1$c setting because the Kagom\'e
plane is $x \approx 0$ in the this setting. The last column, S, shows
the non-zero spin component in the trimer phase and is taken as along
the $a-$, $b-$, and $c-$axis respectively in Figure\ref{FIG8}. For the
N\'eel model, we set the spin magnitude to 0.5 instead of 1/3 and 1/6.
The Cu$_3$ sites are in the triangular lattice planes which
interleave the Kagom\'e planes, but our calculations do not
include their moments.}
\vspace{0.1 in}
\begin{tabular} {||c || c c c | c ||}
\hline \hline
Cu-sites & $x$ & $y$ & $z$  & S \\ 
\hline
Cu$_1$(1) & 0 & 0 & 0 & -1/6   \\
Cu$_1$(2) & 0 & 1/2 & 1/2 & 1/6  \\
\hline
Cu$_2$(2) &\  0.0072\ &\ 0.2658\ &\ 0.2409\  &\ -1/3 \ \\
Cu$_2$(3) &\ 0.9929\ & \ 0.7658\ &\ 0.2591\   & -1/3 \\
Cu$_2$(4) &\ 0.9929\ & \ 0.7342\ &\ 0.7591\ & -1/3 \\
Cu$_2$(2) &\ 0.0072\ & \ 0.2342\ &\ 0.7409\  &\ 1/3 \\
\hline
Cu$_3$(1) & 1/2 & 0 & 1/2  & 0  \\
Cu$_3$(2) & 1/2 & 1/2 & 0  & 0 \\
\hline
\end{tabular}
\end{table}

\section{INELASTIC SCATTERING SCATTERING CROSS SECTION}

In this section we evaluate the inelastic cross section for
the anti configuration at zero temperature.
% with the same caveats asmentioned above Eq. (\ref{SWG}).  
To do
this we will construct the appropriate response functions, namely
\begin{eqnarray}
\langle \langle A; B \rangle \rangle & \equiv & \sum_n \langle 0 |
A| n \rangle \langle n | B | 0 \rangle \delta( E_n - \hbar \omega) \ ,
\label{ABEQ} \end{eqnarray}
where $|0\rangle$ denotes the ground state and the sum is over all
states $|n\rangle$ with energy $E_n$ relative to the ground state.
Here the operators $A$ and $B$ are proportional to the
Fourier transforms of the spin operators. In particular we will need
\begin{eqnarray}
{\cal S}_{\alpha \beta} ({\bf q}, \omega ) &=& 
\langle \langle S_\alpha ({\bf q}) ; S_\beta (-{\bf q}) \rangle \rangle \ .
\end{eqnarray}
Thus
\begin{eqnarray}
{\cal S}_{+-}({\bf q},\omega ) &=& \sum_n | \langle n | S_-({\bf q})| 0
\rangle|^2 \delta(E_n - \hbar \omega ) \ , \nonumber \\
{\cal S}_{-+}({\bf q}, \omega ) &=& \sum_n | \langle n | S_+({\bf q})| 0
\rangle|^2 \delta(E_n - \hbar \omega ) \ , \nonumber \\
{\cal S}_{zz}({\bf q},\omega ) &=& \sum_n | \langle n | S_z({\bf q})| 
0 \rangle|^2 \delta ( E_n - \hbar \omega ) \ .
\label{GFEQ} \end{eqnarray}
(We later set $\hbar =1$.) To analyze the single-magnon contributions to
these quantities we need to relate the spin operators to the
normal mode operators.  Note that when we evaluate Eq. (\ref{GFEQ})
at zero temperature, only contributions to the operator $S_\beta({\bf q})$
proportional to creation operators are nonzero.  We will quote results for
the transverse and longitudinal cross sections, given respectively by
\begin{eqnarray}
I_{\rm trans}({\bf q}, \omega ) &=& 
{\cal S}_{+-}({\bf q}, \omega) + {\cal S}_{-+}({\bf q}, \omega ) , \nonumber \\
I_{\rm long} ({\bf q}, \omega) &=& {\cal S}_{zz}({\bf q}, \omega ) \ .
\end{eqnarray}
In the calculations which follow we use the notation introduced in Sec. VI.

\subsubsection{GROUND-STATE EXCITATIONS}

We first consider inelastic scattering from pseudo-spin waves.  
Accordingly, we discuss spin-wave theory for this
situation.  We express the pseudo-spin operators in terms of boson
creation operators, $c_A^\dagger$ and $c_B^\dagger$ for the A (up)
and B (down) trimers, respectively, as
\begin{eqnarray}
\sigma_z(A) &=& 1/2 - c_A^\dagger c_A \nonumber \\
\sigma_z (B) &=& -1/2 +  c_A^\dagger c_A
\end{eqnarray}
and  (with $\sigma_\pm = \sigma_x \pm i \sigma_y$)
\begin{eqnarray}
\sigma_-(A) &=& c_A^\dagger \ , \ \ \  \sigma_-(B) = c_B \ .
\end{eqnarray}
Then, following the standard spin-wave treatment for such a spin
1/2 system we write
\begin{eqnarray}
c_A({\bf q}) &=& N^{-1/2}  \sum_{\bf R \in A} e^{-i {\bf q} \cdot {\bf R}}
c_A({\bf R})\ ,
\end{eqnarray}
and similarly for $c_B({\bf q})$, where ${\bf R}$ is summed over all the $N$
positions of A trimers.  Then the boson Hamiltonian 
${\cal H} \equiv \sum_{\bf q} {\cal H}_{\bf q}$ at quadratic order is
\begin{eqnarray}
{\cal H}_{\bf q} &=& 2 {\cal J} \{ c_A^\dagger({\bf q}) c_A ({\bf q})
+ c_B^\dagger(-{\bf q}) c_B(-{\bf q}) \nonumber \\
&& \hspace*{-0.2 in} + \gamma({\bf q})
[c_A^\dagger({\bf q}) c_B^\dagger(-{\bf q}) + c_A({\bf q}) c_B(-{\bf q})]\} \ .
\end{eqnarray}
Then the operators which create normal modes are $\rho^\dagger({\bf q})$
and $\eta^\dagger({\bf q})$, which are determined by
\begin{eqnarray}
c_A^\dagger({\bf q}) &=& l({\bf q}) \rho^\dagger({\bf q})
- m({\bf q}) \eta(-{\bf q})
\end{eqnarray}
and
\begin{eqnarray}
c_B(-{\bf q}) &=& - m({\bf q}) \rho^\dagger({\bf q})
+ l({\bf q}) \eta(-{\bf q}) \ ,
\end{eqnarray}
where
\begin{eqnarray}
l({\bf q})^2 &=& \frac{1 + \epsilon({\bf q})}{2 \epsilon({\bf q})}\ , \ \ \ \
m({\bf q})^2 = \frac{1 - \epsilon({\bf q})}{2 \epsilon({\bf q})}\ ,
\nonumber \\ l({\bf q})m({\bf q})
&=& \frac{\gamma({\bf q})}{2 \epsilon({\bf q})} \ ,
\end{eqnarray}
with $\epsilon({\bf q}) = [ 1 - \gamma({\bf q})^2]^{1/2}$.
Apart from the constant zero point energy one has 
\begin{eqnarray}
{\cal H} = \sum_{\bf q} \omega ({\bf q}) [ \rho^\dagger({\bf q}) \rho ({\bf q})
+ \eta^\dagger({\bf q}) \eta ({\bf q}) ] \ ,
\end{eqnarray}
where Eq. (\ref{EQAF}) is $\omega ({\bf q}) =  2 {\cal J} \epsilon({\bf q})$.

Using Eq. (\ref{WIGECK}) we note that the Fourier transform of the
spin operators is
\begin{eqnarray}
S_\alpha ({\bf q}) &=& N^{-1/2} \sum_{{\bf R} \in A} \sigma_\alpha ({\bf R})
e^{-i {\bf q} \cdot {\bf R}} \tau_A({\bf q}) \nonumber \\ &&
\ + N^{-1/2} \sum_{{\bf R} \in B} \sigma_\alpha ({\bf R})
e^{-i {\bf q} \cdot {\bf R}} \tau_B({\bf q}) \ .
\end{eqnarray}
Here we have introduced the trimer form factors
\begin{eqnarray}
\tau_X({\bf q}) = \frac{4}{3} \cos ( {\bf q} \cdot \hat n_X) -
\frac{1}{3} \ ,
\label{TAU1} \end{eqnarray}
where $\hat n_X$ incorporates the locations of the sites
of trimer $X$ relative to its center of gravity:
\begin{eqnarray}
\hat n_A  &=& s(0,1/2,\sqrt3 /2) \nonumber \\
\hat n_B  &=& s(0,-1/2,\sqrt3 /2) \ .
\end{eqnarray}
Within the approximation of a N\'eel state
\begin{eqnarray}
\tau_X({\bf q}) = 2 \cos ( {\bf q} \cdot \hat n_X) - 1 \ .
\label{TAU2} \end{eqnarray}
When $B$ in Eq. (\ref{ABEQ}) is proportional to $S_- ({\bf q})$ we have
(at zero temperature)
\begin{eqnarray}
B &=&  \tau_A({\bf q}) c_A^\dagger({\bf q}) + \tau_B({\bf q}) c_B(-{\bf q}) 
\nonumber \\ & \rightarrow& 
[ l({\bf q}) \tau_A({\bf q}) - \tau_B({\bf q}) m({\bf q}) ]
\rho^\dagger({\bf q}) + \dots
\end{eqnarray}
where the dots indiocate terms involving $\eta({\bf q})$ which do not
contribute at zero temperature.  In $I_{\rm trans}$ 
we also have the contribution when $B$ in Eq.
(\ref{ABEQ}) is proportional to $S_+({\bf q})$, in which case
\begin{eqnarray}
B &=& [ l({\bf q}) \tau_B({\bf q}) - \tau_A({\bf q}) m({\bf q}) ]
\eta^\dagger(- {\bf q}) + \dots 
\end{eqnarray}
Thus the contribution to the inelastic transverse cross section is given by
\begin{eqnarray}
{\cal I}_{\rm trans} ({\bf q}, \omega)
&=& \{ [l({\bf q})^2 + m({\bf q})^2][\tau_A({\bf q})^2 
+ \tau_B({\bf q})^2 ] \nonumber \\ && - 4 l({\bf q})m({\bf q}) \tau_A({\bf q}) 
\tau_B({\bf q}) \} \delta[\omega- \omega({\bf q})]
\nonumber \\ &=& \frac{1}{\epsilon({\bf q})} \{
[ \tau_A({\bf q})^2 + \tau_B({\bf q})^2 ] \nonumber \\ && \ - 2 \gamma({\bf q})
\tau_A({\bf q}) \tau_B({\bf q}) \} \delta[\omega- \omega({\bf q})] \ .
\end{eqnarray}
In the case of a standard two-sublattice antiferromagnet, one has the
same result but with 
\newline
$\tau_A({\bf q})= \tau_B({\bf q})=1$. In that case
inelastic scattering cross section for spin waves alternates in
intensity as one goes from one Brillouin zone to the next due
to the alternating sign of $\gamma({\bf q})$.
Here the result is more complicated because of the form factor of
the unit cell, reflected by the factor $\tau_X({\bf q})$. 

%\begin{table}
%\caption{Inelastic intensities at wave vector $(0,K,L)$ in
%reciprocal lattice units (where the wave vector in real units is
%$K(\pi/s)\hat y + L[\pi/ (\sqrt 3 s)] \hat z$. In the column headed 
%AF (T) we show the result for the two sublattice antiferromagnet
%(the trimer model of the present paper).}
%
%\vspace{0.2 in}
%\begin{tabular} {| c c c c |} \hline
%\ \ \ $K$ \ \ \ &\ \ \  $L$\ \ \  &\ \ \  AF\ \ \  &\ \ \ T\ \ \ \\
%even & odd & $\infty$ & $\infty$ \\
%odd & even & $\infty$ & $\infty$ \\
%odd & odd & $0$ & $\infty$ \\
%even & even & $0$ & $0$ \\
%\hline
%\end{tabular}
%\end{table}

\subsubsection{EXCITONS NEAR ENERGY $J$}

To get the response near energy $J$ for the antiferro case
we need to construct the nonzero matrix elements required
to evaluate Eq. (\ref{GFEQ}).  To obtain the cross section
near energy $J$ we only consider contributions which involve
$a_1^\dagger ({\bf R})$ or $a_2^\dagger ({\bf R})$. From 
Eq. (\ref{EQAA}) and following we see that the only nonzero
contributions of this type are, 
\begin{eqnarray}
S_-(a, {\bf R}) &=& a_1^\dagger ({\bf R}) / \sqrt 3 = -
S_- (c,{\bf R}) \nonumber \\
S_z(a, {\bf R}) &=& a_2^\dagger ({\bf R}) / \sqrt {12} = -
S_z (c,{\bf R}) \ ,
\end{eqnarray}
where ${\bf R}$ is an A, or up, trimer and
\begin{eqnarray}
S_+(a, {\bf R}) &=& a_2^\dagger ({\bf R}) / \sqrt 3 = -
S_+ (c,{\bf R}) \nonumber \\
S_z(a, {\bf R}) &=& -a_1^\dagger ({\bf R}) / \sqrt {12} = -
S_z (c,{\bf R}) \ ,
\end{eqnarray}
when ${\bf R}$ is a B, or down, trimer. These results lead to
\begin{eqnarray}
S_- ({\bf q}) &=& a_{1A}^\dagger ({\bf q}) [e^{i {\bf q} \cdot {\bf n}_A}
- e^{-i {\bf q} \cdot {\bf n}_A}]/ \sqrt 3 \nonumber \\
&=& (2i/\sqrt 3) \xi_A ({\bf q}) a_{1A}^\dagger ({\bf q}) \ ,
\end{eqnarray}
where $\xi_X ({\bf q}) = \sin ({\bf q} \cdot {\bf n}_X )$.  Similarly
\begin{eqnarray}
S_+({\bf q}) &=& (2 i / \sqrt 3 ] \xi_B({\bf q}) 
a_{2B}^\dagger ({\bf q}) \ , \nonumber \\
S_z({\bf q}) &=& (i / \sqrt {3} )
[\xi_A({\bf q}) a_{2A}^\dagger ({\bf q}) 
- \xi_B({\bf q}) a_{1B}^\dagger ({\bf q})]  \ .
\end{eqnarray}
Then, using Eq. (\ref{GFEQ}), we have
\begin{eqnarray}
I_{\rm trans} &=& (4 \xi_A({\bf q})^2/3) \langle \langle a_{1A} ;
a_{1A}^\dagger \rangle \rangle \nonumber \\ && \ + (4 \xi_B({\bf q})^2/3)
\langle \langle a_{2B} ; a_{2B}^\dagger \rangle \rangle \ ,
\end{eqnarray}
where Eq. (\ref{1A2B}) gives
\begin{eqnarray}
\langle \langle a_{1A} ; a_{1A}^\dagger \rangle \rangle &=&
\langle \langle a_{2B} ; a_{2B}^\dagger \rangle \rangle =
\delta [ \omega - \omega_{1A} ({\bf q}) ] \ .
\end{eqnarray}
Also Eq. (\ref{RHOEQ}) gives
\begin{eqnarray}
a_{1B}^\dagger ({\bf q}) &=& [ \rho_+^\dagger ({\bf q}) + 
\rho_-^\dagger ({\bf q})\ / \sqrt 2 \nonumber \\
a_{2A}^\dagger ({\bf q}) &=& [ \rho_+^\dagger ({\bf q}) -
\rho_-^\dagger ({\bf q})\ / \sqrt 2 \ ,
\end{eqnarray}
so that
\begin{eqnarray}
S_z({\bf q})&=& (i / \sqrt 6) \Biggl( [\xi_A({\bf q}) - \xi_B ({\bf q}) ]
\rho_+^\dagger ({\bf q}) \nonumber \\ &&
- [\xi_A({\bf q}) + \xi_B ({\bf q}) ]
\rho_-^\dagger ({\bf q}) ] \Biggr) \ .
\end{eqnarray}
Then we obtain
\begin{eqnarray}
&& \langle \langle S_z({\bf q}) ; S_z(-{\bf q}) \rangle \rangle \nonumber \\
%&& = \langle \langle \xi_A({\bf q}) a_{2,A}({\bf q}) 
%+ \xi_B({\bf q}) a_{1,B}({\bf q}) ; \nonumber \\
%&& \xi_A({\bf q}) a_{2,A}^\dagger({\bf q}) 
%+ \xi_B({\bf q}) a_{1,B}^\dagger({\bf q}) \rangle \rangle \nonumber \\
&& = (1/2) [\xi_A({\bf q}) + \xi_B({\bf q})]^2 \langle \langle
\rho_+({\bf q}) ; \rho_+^\dagger({\bf q}) \rangle \rangle \nonumber \\
&& + (1/2) [\xi_A({\bf q}) - \xi_B({\bf q})]^2 \langle \langle
\rho_-({\bf q}) ; \rho_-^\dagger({\bf q}) \rangle \rangle \ , 
\end{eqnarray}
where Eq. (\ref{PMEQ}) gives that
\begin{eqnarray}
\langle \langle \rho_\pm({\bf q}) ; \rho_\pm^\dagger({\bf q}) \rangle \rangle 
&=& \delta [\omega - \omega_\pm ({\bf q})] \ .
\end{eqnarray}

\subsubsection{EXCITONS NEAR ENERGY $3J/2$}

Here we keep only contributions involving creation operators
$a_n^\dagger$, with $n >2 $. In this case
\begin{eqnarray}
S_-(a,{\bf R}) &=& - a_{4A}^\dagger / \sqrt {18} = S_-(c,{\bf R}) \nonumber \\
S_-(b,{\bf R}) &=& 2 a_{4A}^\dagger / \sqrt{18} \nonumber \\
S_+(a,{\bf R}) &=& a_{6A}^\dagger / \sqrt {6} = S_+(c,{\bf R}) \nonumber \\
S_+(b,{\bf R}) &=& -2 a_{6A}^\dagger / \sqrt{6} \nonumber \\
S_z(a,{\bf R}) &=& - a_{5A}^\dagger / \sqrt {18} = S_-(c,{\bf R}) \nonumber \\
S_z(b,{\bf R}) &=& 2 a_{5A}^\dagger / \sqrt{18} \nonumber \\
\end{eqnarray}
when ${\bf R}$ is an A, up, site.  Also
\begin{eqnarray}
S_+(a,{\bf R}) &=& a_{5B}^\dagger / \sqrt {18} = S_-(c,{\bf R}) \nonumber \\
S_+(b,{\bf R}) &=& -2 a_{5B}^\dagger / \sqrt{18} \nonumber \\
S_-(a,{\bf R}) &=& - a_{3B}^\dagger / \sqrt {6} = S_-(c,{\bf R}) \nonumber \\
S_-(b,{\bf R}) &=& 2 a_{3B}^\dagger / \sqrt{6} \nonumber \\
S_z(a,{\bf R}) &=& - a_{4B}^\dagger / \sqrt {18} = S_-(c,{\bf R}) \nonumber \\
S_z(b,{\bf R}) &=& 2 a_{4B}^\dagger / \sqrt{18} \nonumber \\
\end{eqnarray}
when ${\bf R}$ is an B, down, site.  Thus
\begin{eqnarray}
S_-({\bf q}) &=& - \mu_A({\bf q}) a_{4A}^\dagger({\bf q})/\sqrt{18}
\nonumber \\ && \ 
- \mu_B({\bf q}) a_{3B}^\dagger({\bf q}) /\sqrt{6} \nonumber \\
S_+({\bf q}) &=& \mu_A({\bf q}) a_{6A}^\dagger({\bf q})/ \sqrt {6} \nonumber \\
&& - \mu_B({\bf q}) a_{5B}^\dagger ({\bf q})/\sqrt{18} \nonumber \\
S_z({\bf q}) &=& -[\mu_A({\bf q}) a_{5A}^\dagger ({\bf q})
+ \mu_B({\bf q}) a_{4B}^\dagger ({\bf q}) ] / \sqrt{18} \ ,
\end{eqnarray}
where
\begin{eqnarray}
\mu_X ({\bf q}) = 2 \cos ({\bf q} \cdot {\bf n}_X) -1 \ .
\end{eqnarray}
The intensities can be obtained by inverting the transformation
which diagonalizes $V(3J/2)$ whose eigenvalues are given in Eq. (\ref{MODESJA}).
Since the algebraic expression for the mode intensities are
too complicated to be enlightening, we confine ourselves to some
general remarks.
We verify that $S_-({\bf q})$ involves the third and fourth rows and columns
of the dynamical matrices of Eqs. (\ref{3FA}) and (\ref{3FB}).  Likewise
$S_+({\bf q})$ involves the seventh and eighth rows and columns
of the dynamical matrices of Eqs. (\ref{3FA}) and (\ref{3FB}).  
Thus the transverse response is associated with modes $\omega_4$ and
$\omega_5$, in agreement with our previous identification.  
Similarly we confirm the identification of $\omega_2$ and $\omega_3$
as belonging to the longitudinal response.

\section{CONCLUSIONS}

We may summarize our results as follows.

\noindent
A) The lowest energy modes of the trimer system shown in Fig. \ref{1OM}
are only slightly different from what one gets (see Fig. \ref{NEEL1})
using the N\'eel approximation for the ground state.  There is a slight
difference in symmetry in that the breaking of degeneracy of N\'eel
spin wave in nonspecial directions does not occur in leading order of
perturbation theory within the trimer approximation.

\noindent
B) The elastic diffraction pattern shows differences (see Fig.
\ref{FIG8}) which, in principle, allow one to distinguish 
between a trimer system and one that is closer to the N\'eel limit.

\noindent 
C)  The excitation spectra at high energy
we have obtained show dramatic differences between
the trimer and N\'eel limits.  In the former case, well defined
modes appear in the longitudinal response functions. In general the
trimer limit gives rise to many more elementary excitations and
thereby provides a conclusive way to identify a system as being in the
trimer limit.

\noindent
D)  A possible future project would be to develop an interpolation
scheme to pass between the qualitatively different N\'eel and
trimer limits.

ACKNOWLEDGMENTS.  ABH was supported in part by a grant from
the department of commerce.

\begin{appendix}

\section{Antiferro Excitations at Energy $J$}
%SEARCH A
Here position coordinates are given relative to ${\bf R}$ a
lattice site occupied by an `up' trimer.  Thus $a_2(0)$ denotes
$a_2({\bf R})$, $a_1(\deltav_1)$ denotes $a_1({\bf R}+\deltav_1)$
and so forth.  We treat the interaction of one of the spins (a, b, or c)
of the trimer at ${\bf R}$ with one of the spins (a, b, or c) of
a neighboring trimer at ${\bf R} +\deltav_n$, for $n=1,2,3,4$.
In this section we drop all terms referring to states $n >2$ since
such states occur at energy $3J/2$.  Also, as mentioned, we drop
all terms which are off diagonal in $J$.

\subsection{a at 0 interacts with b at $\deltav_3$}

Within the band at energy $J$ we may write
\begin{eqnarray}
S_-(a) &=& a_1^\dagger(0)/\sqrt 3 \ , \hspace{0.15 in}
S_+(a) = a_1(0)/\sqrt 3 \ , \\
S_z(a) &=& \frac{1}{3} + \frac{a_2^\dagger(0)}{\sqrt{12}} + 
\frac{a_2(0)}{\sqrt{12}} - \frac{n_1(0)}{3} - \frac{n_2(0)}{3}
\end{eqnarray}
and 
\begin{eqnarray}
S_-(b,\deltav_3) &=&0 \ , \hspace{0.4 in} S_+(b,\deltav_3)=0 \\
S_z(b,\deltav_3) &=& \frac{1}{6} - \frac{2n_1(\deltav_3)}{3} + 
\frac{n_2(\deltav_3)}{3} \ .
\end{eqnarray}
Thus this interaction leads to the Hamiltonian
\begin{eqnarray}
{\cal H} &=&  j[ -2 n_1(\deltav_3) + 
n_2(\deltav_3) ]/9 \nonumber \\ &&
- j[ n_1(0) + n_2(0) ]/18 \ .
\end{eqnarray}

\subsection{a at 0 interacts with c at $\deltav_3$ and $\deltav_4$}
%A2
Here
\begin{eqnarray}
S_-(a) &=& a_1^\dagger(0)/\sqrt 3 , \ \
S_+(a) = a_1(0)/\sqrt 3 , \\
S_z(a) &=& \frac{1}{3} + \frac{a_2^\dagger(0)}{\sqrt{12}} 
+ \frac{a_2(0)}{\sqrt{12}} - \frac{n_1(0)}{3} - \frac{n_2(0)}{3}
\end{eqnarray}
and, where $\delta$ assumes the values $\deltav_3$ and $\deltav_4$,
\begin{eqnarray}
S_-(c,\deltav) &=& -a_2(\deltav)/\sqrt 3 , \\ 
S_+(c,\deltav) &=& -a_2^\dagger(\deltav)/\sqrt 3\ , \\
S_z(c,\deltav) &=& -1/3 + a_1^\dagger(\deltav)/\sqrt{12}
+ a_1(\deltav)/\sqrt{12} \nonumber \\ &&
\ + n_1(\deltav)/3 + n_2(\deltav)/3 \ .
\end{eqnarray}
Thus this interaction leads to the Hamiltonian
\begin{eqnarray}
{\cal H} &=& \frac{k}{9} \sum_\deltav
[ n_1(\deltav) + n_2(\deltav) + n_1(0) + n_2(0) ] \nonumber \\
&& + \frac{k}{12} \sum_\deltav [a_2^\dagger (0) a_1(\deltav)
+ a_1^\dagger (\deltav) a_2 (0) ] \ .
\end{eqnarray}

\subsection{b at 0 interacts with a at $\deltav_2$}
%A3
Here
\begin{eqnarray}
S_-(b) &=& 0 \hspace{0.35 in} S_+(b) = 0  \\
S_z(b) &=& -1/6 + 2n_2(0)/3 - n_1(0)/3
\end{eqnarray}
and
\begin{eqnarray}
S_-(a,\deltav_2) &=& \frac{a_2(\deltav_2)}{\sqrt 3}\ ,   \hspace{0.1 in}
S_+(a,\deltav_2) = \frac{a_2^\dagger(\deltav_2)}{\sqrt 3}\ , \\
S_z(a,\deltav_2) &=& - 1/3 - a_1^\dagger(\deltav_2)/\sqrt {12}
- a_1(\deltav_2)/\sqrt{12} \nonumber \\ && 
+ n_1(\deltav_2)/3 + n_2(\deltav_2)/3 \ . 
\end{eqnarray}
These interactions lead to the Hamiltonian
\begin{eqnarray}
{\cal H} &=& -j[ n_2(\deltav_2) + n_1(\deltav_2) ]/18 \nonumber \\ &&
+ j[ n_1(0) -2 n_2(0) ]/9 \ .
\end{eqnarray}

\subsection{b at 0 interacts with c at $\deltav_4$}
%A4
Here
\begin{eqnarray}
S_-(b) &=& 0 \ , \hspace{0.5 in} S_+(b) = 0\ , \\
S_z(b) &=& - 1/6 + 2n_2 (0)/3- n_1(0)/3
\end{eqnarray}
and
\begin{eqnarray}
S_-(c,\deltav_4) &=& - \frac{a_2(\deltav_4)}{\sqrt 3} , \hspace{0.1 in}
S_+(c,\deltav_4) = - \frac{a_2^\dagger(\deltav_4)}{\sqrt 3} \ , \\
S_z(c,\deltav_4) &=& -1/3 + a_1^\dagger(\deltav_4)/\sqrt{12}
+ a_1(\deltav_4)/\sqrt{12} \nonumber \\ &&
+n_1(\deltav_4)/3 + n_2(\deltav_4)/3 \ .
\end{eqnarray}
These interactions lead to the Hamiltonian
\begin{eqnarray}
{\cal H} &=& -j[ n_1(\deltav_4)
+ n_2(\deltav_4) ]/18 \nonumber \\ &&
+ j[ n_1(0) -2  n_2(0) ]/9 \ .
\end{eqnarray}

\subsection{c at 0 interacts with a at $\deltav_1$ and $\deltav_2$}
%A5
Here
\begin{eqnarray}
S_-(c) &=& - a_1^\dagger(0)/\sqrt{3}, \hspace{0.1 in}
S_+(c) = - a_1(0)/\sqrt{3} , \\
S_z(c) &=& \frac{1}{3} - \frac{a_2^\dagger(0)}{\sqrt {12}}
- \frac{a_2(0)}{\sqrt{12}}
- \frac{n_1(0)}{3} - \frac{n_2(0)}{3} \ .
\end{eqnarray}
and, where $\deltav$ assumes the values $\deltav_1$ and $\deltav_2$,
\begin{eqnarray}
S_-(a,\deltav) &=& 0\ , \hspace{0.35 in} S_+(b,\deltav) = 0\ , \\
S_z(a,\deltav) &=& - 1/3 - a_1^\dagger(\deltav)/\sqrt {12}
- a_1(\deltav)/\sqrt {12} \nonumber \\ &&
+ n_1(\deltav)/3 + n_2(\deltav)/3 \ .
\end{eqnarray}
These interactions lead to the Hamiltonian
\begin{eqnarray}
&& {\cal H} = \frac{k}{9} \sum_\deltav [
n_1(\deltav) + n_2(\deltav) + n_1(0) + n_2(0) ] \nonumber \\
&& + \frac{k}{12} \sum_\deltav [ a_2^\dagger (0) a_1(\deltav)
+ a_1^\dagger (\deltav) a_2 (0) ] \ .
\end{eqnarray}

\subsection{c at 0 interacts with b at $\deltav_1$}
%A6
Here
\begin{eqnarray}
S_-(c) &=& -a_1^\dagger(0)/\sqrt 3, \hspace{0.1 in}
S_+(c) = -a_1(0)/\sqrt 3 ,\\
S_z(c) &=& \frac{1}{3} - \frac{a_2^\dagger(0)}{\sqrt {12}} 
- \frac{a_2(0)}{\sqrt{12}} + \frac{n_1(0)}{3} + \frac{n_2(0)}{3}
\end{eqnarray}
and
\begin{eqnarray}
S_-(b,\deltav_1) &=& 0, \hspace{0.15in} S_+(b,\deltav_1) = 0, \nonumber \\
S_z(b, \deltav_1) &=& \frac{1}{6} - \frac{2n_1(\deltav_1)}{3}
+ \frac{n_2(\deltav_1)}{3} \ .
\end{eqnarray}
These lead to the Hamiltonian
\begin{eqnarray}
{\cal H} &=& - j [ n_1(0) + n_2(0) ]/18 \nonumber \\
&& + j[ -2n_1(\deltav_1) + n_2(\deltav_1) ]/9 \ .
\end{eqnarray}

\subsection{Summary}
%A7
Summing all the above contributions we get the Hamiltonian for the
band at energy $J$ for the antiferro configuration
\begin{eqnarray}
{\cal H}(J) &=& \sum_{\bf R} \Biggr( j[n_1({\bf R}) -5 n_2({\bf R})
+ n_2({\bf R}_1) \nonumber \\ &&
- 5n_1({\bf R}_1)]/9 + 4k[n_1({\bf R}) +n_2({\bf R}) \nonumber \\ &&
+ n_1({\bf R}_1) + n_2({\bf R}_1)]/9 \nonumber \\ && \hspace*{-0.3 in} 
+ \sum_\delta k[ a_2^\dagger({\bf R}) a_1({\bf R}+\deltav) \nonumber \\
&& + a_1^\dagger({\bf R} +\deltav) a_2({\bf R}) ]/12 \Biggr) \  ,
\label{EQA7} \end{eqnarray}
where $\deltav$ is summed over the four values shown in Fig. 1.

\section{Antiferro Excitations at Energy $3J/2$}
\subsection{a at 0 interacts with b at $\deltav_3$}
%B1
Here
\begin{eqnarray}
S_-(a) &=& - a_4^\dagger(0)/\sqrt{18} + a_6(0)/\sqrt 6 \ , \nonumber \\
S_+(a) &=& - a_4(0)/\sqrt {18} + a_6^\dagger(0)/\sqrt 6 , \\
S_z(a) &=& \frac{1}{3} - \frac{a_5^\dagger(0)}{\sqrt {18}}
- \frac{a_5(0)}{\sqrt{18}}
+ \frac{n_6(0)}{6} \nonumber \\ && - \frac{n_5(0)}{6} - \frac{n_4(0)}{2} 
- \frac{5n_3(0)}{6} \ .
\end{eqnarray}
and
\begin{eqnarray}
S_-(b,\deltav_3) &=& - 2a_5(\deltav_3)/\sqrt{18}
+ 2a_3^\dagger(\deltav_3)/\sqrt 6 , \\
S_+(b,\deltav_3) &=& - 2a_5^\dagger(\deltav_3)/\sqrt{18}
+ 2a_3(\deltav_3)/\sqrt 6 ,
\\ S_z(b,\deltav_3) &=& \frac{1}{6} + \frac{2a_4^\dagger(\deltav_3)}{\sqrt{18}}
+ \frac{2a_4(\deltav_3)}{\sqrt{18}}\nonumber \\ &&  + \frac{n_6(\deltav_3)}{3}
- \frac{n_4(\deltav_3)}{3} - \frac{2n_3(\deltav_3)}{3} \ .
\end{eqnarray}
These interactions give rise to the Hamiltonian
\begin{eqnarray}
{\cal H} &=& \sqrt 3 j [ -a_4^\dagger(0) a_3(\deltav_3) 
- a_5^\dagger({\deltav_3}) a_6(0) \nonumber \\ &&
- a_3^\dagger(\deltav_3) a_4(0) 
- a_6^\dagger(0)  a_5(\deltav_3) ]/18 \nonumber \\
&& + j [ 4 n_6(\deltav_3) -4 n_4(\deltav_3) -8 n_3(\deltav_3)
+ n_6(0) \nonumber \\ &&  - n_5(0) -3 n_4(0) -5 n_3(0) ]/36 \nonumber \\
&& + j[ -a_5^\dagger (0) a_4(\deltav_3) 
-a_5(0)a_4^\dagger(\deltav_3) ]/9 \ .
\end{eqnarray}

\subsection{a at 0 interacts with c at $\deltav_3$ and $\deltav_4$}
%B2
Here
\begin{eqnarray}
S_-(a) &=& - a_4^\dagger(0)/\sqrt{18} + a_6(0)/\sqrt 6 \ , \\
S_+(a) &=& - a_4(0)/\sqrt{18} + a_6^\dagger(0)/\sqrt 6 \ , \\
S_z(a) &=& \frac{1}{3} - \frac{a_5^\dagger(0)}{\sqrt {18}}
- \frac{a_5(0)}{\sqrt{18}} + \frac{n_6(0)}{6} \nonumber \\ && 
- \frac{n_5(0)}{6} - \frac{n_4(0)}{2} - \frac{5n_3(0)}{6} \ ,
\end{eqnarray}
and, with $\deltav = \deltav_3$ or $\deltav=\deltav_4$, we have
\begin{eqnarray}
S_-(c,\deltav) &=& a_5(\deltav)/\sqrt{18} - a_3^\dagger(\deltav)/\sqrt 6\ ,\\
S_+(c,\deltav) &=& a_5^\dagger(\deltav)/\sqrt{18} - a_3(\deltav)/\sqrt 6 \ , \\
S_z(c,\deltav) &=& -\frac{1}{3} - \frac{a_4^\dagger(\deltav)}{\sqrt {18}}
- \frac{a_4(\deltav)}{\sqrt{18}} + \frac{5n_6(\deltav)}{6} 
\nonumber \\ && + \frac{n_5(\deltav)}{2}
+ \frac{n_4(\deltav)}{6} - \frac{n_3(\deltav)}{6} \ .
\end{eqnarray}
These interactions lead to the Hamiltonian
\begin{eqnarray}
&& {\cal H} = \sum_\deltav \Biggl[
\frac{\sqrt 3 k}{36} \biggl( a_4^\dagger(0) a_3(\deltav)
+ a_5^\dagger(\deltav) a_6(0) \nonumber \\ &&  + a_6^\dagger(0) a_5(\deltav)
+ a_3^\dagger (\deltav) a_4(0) \biggr) \nonumber \\
&& + \frac{k}{18} \biggl( 5 n_6(\deltav) + 3 n_5 (\deltav)
+ n_4 (\deltav) - n_3(\deltav) \nonumber \\ &&
- n_6(0) + n_5(0) + 3 n_4(0) + 5 n_3(0) \biggr) 
\nonumber \\ && + \frac{k}{18} \biggl( a_5^\dagger (0) a_4(\deltav)
+ a_4^\dagger (\deltav) a_5(0) \biggr) \Biggr] \ .
\end{eqnarray}

\subsection{b at 0 interacts with a at $\deltav_2$}
%B3
Here
\begin{eqnarray}
S_-(b) &=& 2 a_4^\dagger(0)/\sqrt{18} - 2 a_6(0)/\sqrt 6\ , \\
S_+(b) &=& 2 a_4(0)/\sqrt{18} - 2 a_6^\dagger(0)/\sqrt 6 \ ,  \\
S_z(b) &=& -\frac{1}{6}  + \frac{2 a_5^\dagger(0)}{\sqrt{18}} 
+ \frac{2 a_5(0)}{\sqrt {18}} \nonumber \\ && + \frac{2n_6(0)}{3}
+ \frac{n_5(0)}{3} - \frac{n_3(0)}{3}
\end{eqnarray}
and,
\begin{eqnarray}
S_-(a,\deltav_2) &=& a_5(\deltav_2)/\sqrt{18}
- a_3^\dagger(\deltav_2)/\sqrt 6 ,  \nonumber \\
S_+(a,\deltav_2) &=& a_5^\dagger(\deltav_2)/\sqrt{18}
- a_3(\deltav_2)/\sqrt 6 , \\
S_z(a,\deltav_2) &=& -\frac{1}{3} - \frac{a_4^\dagger(\deltav_2)}{\sqrt{18}}
- \frac{a_4(\deltav_2)}{\sqrt{18}} + \frac{5n_6(\deltav_2)}{6} 
\nonumber \\ && + \frac{n_5(\deltav_2)}{2} + \frac{n_4(\deltav_2)}{6} 
- \frac{n_3(\deltav_2)}{6} \ .
\end{eqnarray}
Thus the Hamiltonian for this interaction is
\begin{eqnarray}
&&  {\cal H} = \sqrt 3 j [ -a_4^\dagger(0) a_3(\deltav_2) 
- a_5^\dagger({\deltav_2})a_6(0) \nonumber \\ &&  
- a_3^\dagger (\deltav_2) a_4(0) 
- a_6^\dagger(0)  a_5(\deltav_2) ]/18 \nonumber \\
&& + j [ -5 n_6(\deltav_2) -3 n_5(\deltav_2) - n_4(\deltav_2)
+ n_3(\deltav_2) \nonumber \\ &&
- 8 n_6(0) - 4n_5(0) + 4 n_3(0) ]/36 \nonumber \\
&& + j [ -a_5^\dagger(0) a_4(\deltav_2) 
- a_4^\dagger(\deltav_2) a_5(0) ]/9 \ .
\end{eqnarray}

\subsection{b at 0 interacts with c at $\deltav_4$}
%B4
Here
\begin{eqnarray}
S_-(b) &=& 2 a_4^\dagger(0)/\sqrt{18} - 2 a_6(0)/\sqrt 6 \ , \\
S_+(b) &=& 2 a_4(0)/\sqrt{18} - 2 a_6^\dagger(0)/\sqrt 6 \ ,  \\
S_z(b) &=& -\frac{1}{6}  + \frac{2 a_5^\dagger(0)}{\sqrt{18}} 
+ \frac{2 a_5(0)}{\sqrt {18}} + \frac{2n_6(0)}{3} \nonumber \\
&& + \frac{n_5(0)}{3} - \frac{n_3(0)}{3}
\end{eqnarray}
and
\begin{eqnarray}
S_-(c,\deltav_4) &=& a_5(\deltav_4)/\sqrt{18}
- a_3^\dagger(\deltav_4)/\sqrt 6 , \nonumber \\
S_+(c,\deltav_4) &=& a_5^\dagger(\deltav_4)/\sqrt{18}
- a_3(\deltav_4)/\sqrt 6 , \nonumber \\
S_z(c,\deltav_4) &=& -\frac{1}{3} - \frac{a_4^\dagger(\deltav_4)}{\sqrt {18}}
- \frac{a_4(\deltav_4)}{\sqrt{18}} + \frac{5n_6(\deltav_4)}{6}\nonumber \\ &&
+ \frac{n_5(\deltav_4)}{2} + \frac{n_4(\deltav_4)}{6} 
- \frac{n_3(\deltav_4)}{6} \ .
\end{eqnarray}
These results lead to the Hamiltonian
\begin{eqnarray}
&&  {\cal H} = \sqrt 3 j [ -a_4^\dagger (0) a_3(\deltav_4) 
- a_5^\dagger({\deltav_4}) a_6(0)  \nonumber \\ &&
- a_3^\dagger (\deltav_4) a_4(0) 
- a_6^\dagger (0)  a_5(\deltav_4) ]/18 \nonumber \\
&& + j [ -5 n_6(\deltav_4) -3 n_5(\deltav_4) - n_4(\deltav_4)
+ n_3(\deltav_4) \nonumber \\ &&
- 8n_6(0) - 4n_5(0) + 4n_3(0) ]/36 \nonumber \\
&& + j [ -a_5^\dagger (0) a_4(\deltav_4) 
- a_4^\dagger(\deltav_4) a_5(0) ]/9 \ .
\end{eqnarray}

\subsection{c at 0 interacts with a at $\deltav_1$ and $\deltav_2$}
%B5
Here
\begin{eqnarray}
S_-(c) &=& - a_4^\dagger(0)/\sqrt{18} + a_6(0)/\sqrt 6 \ , \\
S_+(c) &=& - a_4(0)/\sqrt {18} + a_6^\dagger(0)/\sqrt 6 \\
S_z(c) &=& \frac{1}{3} - \frac{a_5^\dagger(0)}{\sqrt {18}} 
- \frac{a_5(0)}{\sqrt{18}} + \frac{n_6(0)}{6} \nonumber \\ &&
- \frac{n_5(0)}{6} - \frac{n_4(0)}{2 }-  \frac{5n_3(0)}{6}
\end{eqnarray}
and, where $\deltav$ assumes the values $\deltav_1$ and $\deltav_2$,
\begin{eqnarray}
&& S_-(a,\deltav) = a_5(\deltav)/\sqrt{18}
- a_3^\dagger(\deltav)/\sqrt 6 , \nonumber \\
&& S_+(a,\deltav) = a_5^\dagger(\deltav)/\sqrt{18}
- a_3(\deltav)/\sqrt 6 , \\
&& S_z(a,\deltav) = -\frac{1}{3} -  \frac{a_4^\dagger(\deltav)}{\sqrt{18}}
- \frac{a_4(\deltav)}{\sqrt{18}}\nonumber \\ &&
+ \frac{5n_6(\deltav)}{6} +\frac{n_5(\deltav)}{2}
+ \frac{n_4(\deltav)}{6} - \frac{n_3(\deltav)}{6} \ .
\end{eqnarray}
Thus the Hamiltonian from this interaction is
\begin{eqnarray}
&& {\cal H} = \sum_\deltav \Biggl[
\frac{\sqrt 3 k}{36} \biggl( a_4^\dagger(0) a_3(\deltav)
+ a_6(0) a_5^\dagger(\deltav)  \nonumber \\ &&
+ a_4(0) a_3^\dagger(\deltav) 
+ a_6^\dagger (0) a_5(\deltav) \Biggr) \nonumber \\
&& + \frac{k}{18} \biggl( 5 n_6(\deltav) + 3 n_5 (\deltav)
+ n_4 (\deltav) - n_3(\deltav) \nonumber \\ &&
-2 n_6(0) + 2n_5(0) + 6 n_4(0) + 10 n_3(0) \nonumber \\ &&
+ a_5^\dagger (0) a_4(\deltav)
+ a_5(0) a_4^\dagger (\deltav) \biggr) \Biggr] \ .
\end{eqnarray}

\subsection{c at 0 interacts with b at $\deltav_1$}
%B6
Here
\begin{eqnarray}
S_-(c) &=& - a_4^\dagger(0)/\sqrt{18} + a_6(0)/\sqrt 6  \\
S_+(c) &=& - a_4(0)/\sqrt {18} + a_6^\dagger(0)/\sqrt 6 \\
S_z(c) &=& \frac{1}{3} - \frac{a_5^\dagger(0)}{\sqrt{18}} 
- \frac{a_5(0)}{\sqrt{18}} + \frac{n_6(0)}{6} \nonumber \\ &&
- \frac{n_5(0)}{6} - \frac{n_4(0)}{2} - \frac{5n_3(0)}{6} 
\end{eqnarray}
and
\begin{eqnarray}
S_-(b,\deltav_1) &=& - 2a_5(\deltav_1)/\sqrt{18} 
+ 2a_3^\dagger(\deltav_1)/\sqrt 6 \\
S_+(b,\deltav_1) &=& - 2a_5^\dagger(\deltav_1)/\sqrt{18}
+ 2a_3(\deltav_1)/\sqrt 6 , \\
S_z(b,\deltav_1) &=& \frac{1}{6}  + \frac{2a_4^\dagger(\deltav_1)}{\sqrt{18}}
+ \frac{2a_4(\deltav_1)}{\sqrt{18}} \nonumber \\ && \hspace*{-0.2 in} 
+ \frac{n_6(\deltav_1)}{3} - \frac{n_4(\deltav_1)}{3}
- \frac{2n_3(\deltav_1)}{3} \ .
\end{eqnarray}
Thus the Hamiltonian from this interaction is
\begin{eqnarray}
&&  {\cal H} = \sqrt 3 j [ -a_4^\dagger(0) a_3(\deltav) 
- a_5^\dagger({\deltav}) a_6(0) \nonumber \\ &&  - a_3^\dagger (\deltav) a_4(0)
- a_6^\dagger(0)  a_5(\deltav) ]/18 \nonumber \\
&& + j [ 4 n_6(\deltav) -4 n_4(\deltav) -8 n_3(\deltav)
+ n_6(0) \nonumber \\ && - n_5(0) -3 n_4(0) -5 n_3(0) ]/36 \nonumber \\
&& + j [ -a_5^\dagger (0) a_4(\deltav) 
- a_4^\dagger(\deltav) a_5(0) ]/9 \ .
\end{eqnarray}

\subsection{Summary}
%B7

Summing all the above contributions we get the Hamiltonian for the
band at energy $3J/2$ for the antiferro configuration
\begin{eqnarray}
{\cal H} &=& j \sum_{\bf R} \Biggl[ [ - n_3({\bf R}) -3 n_4({\bf R})
-5 n_5({\bf R}) - 7 n_6({\bf R}) \nonumber \\
&&- n_6({\bf R}_1) - 3 n_5 ({\bf R}_1) - 5 n_4({\bf R}_1)
- 7 n_3({\bf R}_1)]/18 \nonumber \\ &&
+ \sum_\deltav \Biggl( -[a_5^\dagger ({\bf R}) a_4({\bf R}+\deltav)
+ a_4^\dagger({\bf R}+\deltav) a_5({\bf R})]/9 \nonumber \\ &&
- \sqrt 3 [ a_4^\dagger({\bf R}) a_3({\bf R}+\deltav)
+ a_3^\dagger ({\bf R}+\deltav) a_4({\bf R}) \nonumber \\ &&
+ a_5^\dagger ({\bf R}+ \deltav) a_6({\bf R})
+ a_6^\dagger({\bf R}) a_5({\bf R}+\deltav) ]/18 \Biggr) \Biggr]\nonumber \\ &&
+ k \sum_{\bf R} \Biggl[ 2 [-n_6({\bf R}) + n_5({\bf R}) + 3 n_4({\bf R})
+ 5 n_3({\bf R}) \nonumber \\ &&
+5 n_6({\bf R}_1) + 3n_5({\bf R}_1) + n_4({\bf R}_1) - n_3({\bf R}_1)]/9
\nonumber \\ &&
+ \sum_\deltav \Biggl( \sqrt 3 [a_4^\dagger ({\bf R}) a_3 ({\bf R}+\deltav)
+ a_3^\dagger ({\bf R}+\deltav) a_4 ({\bf R}) \nonumber \\ &&
+ a_5^\dagger ({\bf R}+\deltav) a_6 ({\bf R}) 
+ a_6^\dagger ({\bf R}) a_5 ({\bf R}+\deltav)]/36 \nonumber \\ &&
+ [a_5^\dagger ({\bf R}) a_4 ({\bf R}+\deltav)
+ a_4^\dagger ({\bf R}+\deltav) a_5 ({\bf R})]/18 \Biggr) \Biggr] \ .
\label{EQB7} \end{eqnarray}

\section{Ferro Excitations at Energy $J$}

\subsection{a at 0 interacts with b at $\deltav_3$}
%C1
Here
\begin{eqnarray}
S_-(a) &=& a_1^\dagger(0)/\sqrt3 , S_+(a) = a_1(0)/\sqrt 3  , \\
S_z(a) &=& \frac{1}{3} + \frac{a_2^\dagger(0)}{\sqrt{12}}
+ \frac{a_2(0)}{\sqrt{12}} - \frac{n_1(0)}{3} - \frac{n_2(0)}{3} \ ,
\end{eqnarray}
and
\begin{eqnarray}
S_-(b,\deltav_3) &=& 0 \ , \hspace{0.15 in} S_+(b,\deltav_3) = 0\ , \\
S_z(b,\deltav_3) &=& -\frac{1}{6} - \frac{n_1(\deltav_3)}{3}
+ \frac{2n_2(\deltav_3)}{3} \ .
\end{eqnarray}
These results lead to the Hamiltonian
\begin{eqnarray}
{\cal H} &=&  j[ 2 n_2(\deltav_3) - n_1(\deltav_3) ]/9 \nonumber \\
&&  + j[ n_1(0) + n_2 (0)]/18 \ .
\end{eqnarray}

\subsection{a at 0 interacts with c  at $\deltav_3$ and $\deltav_4$}
%C2
Here
\begin{eqnarray}
S_-(a) &=& a_1^\dagger(0)/\sqrt3  , \hspace{0.15 in}
S_+(a) = a_1(0)/\sqrt 3  , \\
S_z(a) &=& \frac{1}{3} + \frac{a_2^\dagger(0)}{\sqrt{12}} 
+ \frac{a_2(0)}{\sqrt{12}} - \frac{n_1(0)}{3} - \frac{n_2(0)}{3}
\end{eqnarray}
and, where $\deltav$ assumes the values $\deltav_3$ and $\deltav_4$,
\begin{eqnarray}
S_-(c,\deltav) &=& - \frac{a_1^\dagger(\deltav)}{\sqrt 3} , \hspace{0.1in}
S_+(c,\deltav) = - \frac{a_1(\deltav)}{\sqrt 3} \ ,  \\
S_z(c,\deltav) &=& 1/3 - a_2^\dagger(\deltav)/\sqrt {12}
- a_2(\deltav)/\sqrt{12} \nonumber \\ && - n_1(\deltav)/3 - n_2(\deltav)/3 \ .
\end{eqnarray}
Thus, we have that
\begin{eqnarray}
&& {\cal H} = \sum_\deltav \Biggl[
- \frac{k}{9} \biggl( n_1(\deltav) + n_2(\deltav) + n_1(0) + n_2(0) \biggr)
\nonumber \\ && + \frac{k}{12} \biggl( - a_2^\dagger (\deltav) a_2(0)
 - a_2^\dagger (\deltav) a_2(0) \biggr) \nonumber \\
&& - \frac{k}{6} \biggl( a_1^\dagger(0) a_1(\deltav)
+ a_1^\dagger (\deltav) a_1(0) \biggr) \Biggr] \ .
\end{eqnarray}

\subsection{b at 0 interacts with a  at $\deltav_2$}
%C3
Here
\begin{eqnarray}
S_-(b) &=& 0\ , \hspace{0.15 in} S_+(b) = 0\ ,  \\
S_z(b) &=& -\frac{1}{6} + \frac{2n_2(0)}{3} - \frac{n_1(0)}{3} \ ,
\end{eqnarray}
and
\begin{eqnarray}
S_-(a,\deltav_2) &=& \frac{a_1^\dagger(\deltav_2)}{\sqrt 3} \ ,
\hspace{0.1 in}
S_+(a,\deltav_2) = \frac{a_1(\deltav_2)}{\sqrt 3} \ , \\
S_z(a,\deltav_2) &=& 1/3 + a_2^\dagger(\deltav_2)/\sqrt{12}
+ a_2(\deltav_2)/\sqrt{12} \nonumber \\ && - n_1(\deltav_2)/3 
- n_2(\deltav_2)/3 \ .
\end{eqnarray}
Thus, we obtain the Hamiltonian
\begin{eqnarray}
&& {\cal H} = j [ n_1(\deltav_2) + n_2(\deltav_2) ]/18 \nonumber \\
&& + j [ - n_1(0) + 2 n_2(0) ]/9 \ .
\end{eqnarray}

\subsection{b at 0 interacts with c  at $\deltav_4$}
%C4
Here
\begin{eqnarray}
S_-(b) &=& 0\ , \hspace{0.15 in} S_+(b) = 0 , \\
S_z(b) &=& - \frac{1}{6} + \frac{2n_2(0)}{3} - \frac{n_1(0)}{3}
\end{eqnarray}
and
\begin{eqnarray}
S_-(c,\deltav_4) &=& - \frac{a_1^\dagger(\deltav_4)}{\sqrt 3} ,
S_+(c,\deltav_4) = - \frac{a_1(\deltav_4)}{\sqrt 3} , \\
S_z(c,\deltav_4) &=& 1/3 - a_2^\dagger(\deltav_4)/\sqrt{12}
- a_2(\deltav_4)/\sqrt{12} \nonumber \\ &&
 -n_1(\deltav_4)/3 -  n_2(\deltav_4)/3 \ .
\end{eqnarray}
These interactions give rise to the Hamltonian
\begin{eqnarray}
&& {\cal H} = j [ n_1(\deltav_4) + n_2(\deltav_4) ]/18 \nonumber \\ &&
+ j [ -n_1(0) +2  n_2(0) ] \ 9 \ .
\end{eqnarray}

\subsection{c at 0 interacts with a  at $\deltav_1$ and $\deltav_2$}
%C5
Here
\begin{eqnarray}
S_-(c) &=& - \frac{a_1^\dagger(0)}{\sqrt{3}} \ , \hspace{0.15 in}
S_+(c) = - \frac{a_1(0)}{\sqrt{3}} \ , \\
S_z(c) &=& 1/3 - a_2^\dagger(0)/\sqrt{12} - a_2(0)/\sqrt{12} \nonumber \\ &&
- n_1(0)/3  - n_2(0)/3 \ . 
\end{eqnarray}
and, where $\deltav$ assumes the values $\deltav_1$ and $\deltav_2$,
\begin{eqnarray}
S_-(a,\deltav) &=& \frac{a_1^\dagger)(\deltav)}{\sqrt 3} \ , \hspace{0.15 in}
S_+(a,\deltav) = \frac{a_1(\deltav)}{\sqrt 3} \ , \\
S_z(a,\deltav) &=& 1/3 + a_2^\dagger(\deltav)/\sqrt {12}
+ a_2(\deltav)/\sqrt{12}  \nonumber \\ && -n_1 (\deltav)/3
- n_2(\deltav)/3 \ .
\end{eqnarray}
These results lead to the Hamiltonian
\begin{eqnarray}
&& {\cal H} = \sum_\deltav \Biggl[
- \frac{k}{9} \biggl( n_1(\deltav) + n_2(\deltav) + n_1(0) + n_2(0) \biggr)
\nonumber \\ && - \frac{k}{12} \biggl(
a_2^\dagger(0) a_2(\deltav) + a_2^\dagger (\deltav) a_2(0) \biggr)
\nonumber \\ && - \frac{k}{6} \biggl( a_1^\dagger(0) a_1(\deltav)
+ a_1^\dagger(\deltav) a_1(0)
\biggr) \Biggr] \ .
\end{eqnarray}

\subsection{c at 0 interacts with b  at $\deltav_1$}
%C6
Here
\begin{eqnarray}
S_-(c) &=& - \frac{a_1^\dagger(0)}{\sqrt{3}}\ , \hspace{0.1 in}
S_+(c) = - \frac{a_1(0)}{\sqrt{3}} \ , \\
S_z(c) &=& \frac{1}{3} - \frac{a_2^\dagger(0)}{\sqrt{12}}
- \frac{a_2(0)}{\sqrt{12}} - \frac{n_1(0)}{3} - \frac{n_2(0)}{3} \ ,
\end{eqnarray}
and
\begin{eqnarray}
S_-(b,\deltav_1) &=& 0 \ , \hspace{0.1 in} S_+(b,\deltav_1) = 0 \\
S_z(b,\deltav_1) &=& -\frac{1}{6} + \frac{2n_2(\deltav_1)}{3}
- \frac{n_1(\deltav_1)}{3}
\end{eqnarray}
These terms give rise to the Hamiltonian
\begin{eqnarray}
&& {\cal H} = j[ n_1(0) + n_2(0) ]/18 \nonumber \\ &&
+ j [ 2n_2(\deltav_1) - n_1(\deltav_1) ]/9 \ .
\end{eqnarray}

\subsection{Summary}
%C7
Summing all the above contributions we get the Hamiltonian for the
band at energy $J$ for the ferroconfiguration as
\begin{eqnarray}
{\cal H } &=& \sum_{\bf R} \Biggl[ j[5n_2({\bf R})- n_1({\bf R})
- n_1({\bf R}_1) + 5n_2({\bf R}_1) \nonumber \\ &&
- 4k [ n_1({\bf R}) + n_2({\bf R}) + n_1({\bf R}_1) + n_2({\bf R}_1) ] /9
\nonumber \\ && \hspace*{-0.2 in} - \frac{k}{12}
\sum_\deltav \Biggr( [2a_1^\dagger ({\bf R}) a_1({\bf R}+\deltav)
+a_2^\dagger ({\bf R}) a_2({\bf R}+\deltav) \nonumber \\ &&
+2a_1^\dagger ({\bf R}+\deltav) a_1({\bf R})
+a_2^\dagger ({\bf R}+\deltav) a_2({\bf R}) \Biggr) \Biggr] \ .
\label{EQC7} \end{eqnarray}

\section{Ferro Excitations at Energy $3J/2$}
%D1
\subsection{a at 0 interacts with b at $\deltav_3$}
Here
\begin{eqnarray}
S_-(a) &=& - a_4^\dagger(0)/\sqrt{18} + a_6(0)/\sqrt6\ , \\
S_+(a) &=& - a_4(0)/\sqrt {18} + a_6^\dagger(0)/\sqrt 6 \ , \\
S_z(a) &=& \frac{1}{3} - \frac{a_5^\dagger(0)}{\sqrt{18}}
- \frac{a_5(0)}{\sqrt{18}} + \frac{n_6(0)}{6} \nonumber \\
&&  - \frac{n_5(0)}{6} -\frac{n_4(0)}{2} - \frac{5 n_3(0)}{6} \ ,
\end{eqnarray}
and
\begin{eqnarray}
S_-(b,\deltav_3) &=&  2a_4^\dagger(\deltav_3)/\sqrt{18} 
- 2a_6(\deltav_3)/\sqrt6 \ , \\
S_+(b,\deltav_3) &=& 2a_4(\deltav_3)/\sqrt{18} 
- 2a_6^\dagger(\deltav_3)/\sqrt 6 \ , \\
S_z(b,\deltav) &=& - 1/6 + 2a_5^\dagger(\deltav_3)/\sqrt {18}
+ 2a_5(\deltav_3)/\sqrt{18} \nonumber \\ && \hspace*{-0.2 in}
+ 2n_6(\deltav_3)/3
+ n_5(\deltav_3)/3 - n_3(\deltav_3)/3 \ .
\end{eqnarray}
Thus these interactions lead to the Hamiltonian
\begin{eqnarray}
&&{\cal H} = j [ -a_4^\dagger(0) a_4(\deltav_3) 
- 3 a_6^\dagger({\deltav_3}) a_6(0) \nonumber \\ &&
- a_4^\dagger (\deltav_3) a_4(0)
- 3a_6^\dagger (0)  a_6(\deltav_3) ]/18 \nonumber \\
&& + j [ -4 n_3(\deltav_3) +4 n_5(\deltav_3) + 8 n_6(\deltav_3)
\nonumber \\ && - n_6(0) + n_5(0) + 3 n_4(0) + 5 n_3(0) ]/36 \nonumber \\
&& + j [ -a_5^\dagger (0) a_5(\deltav_3) 
- a_5^\dagger(\deltav_3) a_5(0) ]/9 \ .
\end{eqnarray}

\subsection{a at 0 interacts with c at $\deltav_3$ and $\deltav_4$}
%SECD2
Here
\begin{eqnarray}
S_-(a) &=& - a_4^\dagger(0)/\sqrt{18} + a_6(0)/\sqrt 6\ , \\
S_+(a) &=& - a_4(0)/\sqrt{18} + a_6^\dagger(0)/\sqrt 6 \ ,  \\
S_z(a) &=& \frac{1}{3} - \frac{a_5^\dagger(0)}{\sqrt{18}}
- \frac{a_5(0)}{\sqrt{18}} + \frac{n_6(0)}{6} \nonumber \\ &&
- \frac{n_5(0)}{6} - \frac{n_4(0)}{2}  - \frac{5n_3(0)}{6}
\end{eqnarray}
and, where $\deltav$ assumes the values $\deltav_3$ and $\deltav_4$,
\begin{eqnarray}
S_-(c,\deltav) &=& - a_4^\dagger(\deltav)/\sqrt{18}
+ a_6(\deltav)/\sqrt6 \ , \\ 
S_+(c,\deltav) &=& -a_4(\deltav)/\sqrt{18} 
+ a_6^\dagger(\deltav)/\sqrt 6 \ ,  \\
S_z(c,\deltav) &=& \frac{1}{3} - \frac{a_5^\dagger(\deltav)}{\sqrt {18}}
- \frac{a_5(\deltav)}{\sqrt{18}} - \frac{5n_3(\deltav)}{6}
\\ && - \frac{n_4(\deltav)}{2}  - \frac{n_5(\deltav)}{6} 
+ \frac{n_6(\deltav)}{6} \ .
\end{eqnarray}
Thus
\begin{eqnarray}
&& {\cal H} = \sum_\deltav \Biggl[ \frac{k}{36} \biggl(
a_4^\dagger (\deltav) a_4(0) + 3 a_6^\dagger(\deltav) a_6(0)
\nonumber \\ && + a_4^\dagger(0)  a_4(\deltav)
+ 3a_6^\dagger (0) a_6(\deltav) \biggr) \nonumber \\
&& + \frac{k}{18} \biggl( -5 n_3(\deltav) - 3 n_4 (\deltav)
- n_5 (\deltav) + n_6(\deltav) \nonumber \\ &&
+ n_6(0) - n_5(0) - 3 n_4(0) - 5 n_3(0) \biggr )
\nonumber \\ && + \frac{k}{18} \biggl( a_5^\dagger (0) a_5(\deltav)
+ a_5^\dagger (\deltav_3) a_5(0) \biggr) \Biggr] \ .
\end{eqnarray}

\subsection{b at 0 interacts with a at $\deltav_2$}
%SECD3
\begin{eqnarray}
S_-(b) &=& 2 a_4^\dagger(0)/\sqrt{18} - 2 a_6(0)/\sqrt 6 \ , \\ 
S_+(b) &=& 2 a_4(0)/\sqrt{18} - 2 a_6^\dagger(0)\sqrt 6 \ ,  \\
S_z(b) &=& - 1/6 + 2 a_5^\dagger(0)/\sqrt {18} + 2 a_5(0)/\sqrt {18}
\nonumber \\ &&
+ 2n_6(0)/3 + n_5(0)/3 - n_3(0)/3 \ ,
\end{eqnarray}
and
\begin{eqnarray}
S_-(a,\deltav_2) &=& - a_4^\dagger(\deltav_2)/\sqrt{18}
+ a_6(\deltav_2)/\sqrt 6\ ,  \\
S_+(a,\deltav_2) &=& - a_4(\deltav_2)/\sqrt{18} 
+ a_6^\dagger(\deltav_2)/\sqrt 6 \ , \\
S_z(a,\deltav_2) &=& \frac{1}{3} - \frac{a_5^\dagger(\deltav_2)}{\sqrt{18}}
- \frac{a_5(\deltav_2)}{\sqrt{18}} - \frac{5n_3(\deltav_2)}{6}
\nonumber \\ &&  \hspace*{-0.2 in} - \frac{n_4(\deltav_2)}{2}
- \frac{n_5(\deltav_2)}{6} + \frac{n_6(\deltav_2)}{6} \ .
\end{eqnarray}
Thus, we obtain the Hamiltonian
\begin{eqnarray}
&&  {\cal H} = j [ -a_4^\dagger(0) a_4(\deltav_2) 
- 3 a_6^\dagger({\deltav_2}) a_6(0) \nonumber \\ &&
- a_4^\dagger (\deltav_2) a_4(0) - 3a_6^\dagger(0)  a_6(\deltav_2) \biggr]/18
\nonumber \\ &&
+ j [ 5 n_3(\deltav_2) + 3 n_4(\deltav_2) + n_5(\deltav_2)
\nonumber \\ && - n_6(\deltav_2) + 8 n_6(0) + 4n_5(0) - 4 n_3(0) ]/36 
\nonumber \\ &&
+ j [ -a_5^\dagger(0) a_5(\deltav_2) 
- a_5^\dagger(\deltav_2) a_5(0) ]/9 \ .
\end{eqnarray}

\subsection{b at 0 interacts with c at $\deltav_4$}
%SECD4

Here
\begin{eqnarray}
S_-(b) &=& 2 a_4^\dagger(0)/\sqrt{18} - 2a_6(0)/\sqrt 6 \\
S_+(b) &=& 2 a_4(0)/\sqrt{18} - 2 a_6^\dagger(0)/\sqrt 6  \\
S_z(b) &=& -1/6 + 2 a_5^\dagger(0)/\sqrt{18}
+ 2 a_5(0)/\sqrt {18} \nonumber \\ && + 2n_6(0)/3
+ n_5(0)/3 - n_3(0)/3
\end{eqnarray}
and
\begin{eqnarray}
S_-(c,\deltav_4) &=& - a_4^\dagger(\deltav_4)/\sqrt{18}
+ a_6(\deltav_4)/\sqrt 6 \ , \\
S_+(c,\deltav_4) &=& - a_4(\deltav_4)/\sqrt{18}
+ a_6^\dagger(\deltav_4)/\sqrt 6 \ ,  \\
S_z(c,\deltav_4) &=& \frac{1}{3} - \frac{a_5^\dagger(\deltav_4)}{\sqrt{18}}
- \frac{a_5(\deltav_4)}{\sqrt{18}} - \frac{5n_3(\deltav_4)}{6} \nonumber \\ &&
\hspace*{-0.2 in} - \frac{n_4(\deltav_4)}{2} - \frac{n_5(\deltav_4)}{6} 
+ \frac{n_6(\deltav_4)}{6} \ .
\end{eqnarray}
Thus we obtain the Hamiltonian
\begin{eqnarray}
&&  {\cal H} = j[ -a_4^\dagger (0) a_4(\deltav_4) 
- 3 a_6^\dagger({\deltav_4}) a_6(0) \nonumber \\ &&
- a_4^\dagger (\deltav_4) a_4(0)
- 3a_6^\dagger(0)  a_6(\deltav_4) ]/18 \nonumber \\
&& + j [ 5 n_3(\deltav_4) + 3 n_4(\deltav_4) + n_5(\deltav_4)
- n_6(\deltav_4) \nonumber \\ && + 8n_6(0) + 4n_5(0) - 4n_3(0) ]/36 \nonumber \\
&& + j[ -a_5^\dagger(0) a_5(\deltav_4) -a_5^\dagger(\deltav_4) a_5(0) ]/9 \ .
\end{eqnarray}

\subsection{c at 0 interacts with a at $\deltav_1$ and $\deltav_2$}
%SECD5

Here
\begin{eqnarray}
S_-(c) &=& - a_4^\dagger(0)/\sqrt {18} + a_6(0)/\sqrt 6  \ , \\
S_+(c) &=& - a_4(0)/\sqrt{18} + a_6^\dagger(0)/\sqrt 6 \\
S_z(c) &=& \frac{1}{3} - \frac{a_5^\dagger(0)}{\sqrt {18}}
- \frac{a_5(0)}{\sqrt {18}} + \frac{n_6(0)}{6} 
- \frac{n_5(0)}{6} \nonumber \\ &&
- \frac{n_4(0)}{2} - \frac{5n_3(0)}{6}
\end{eqnarray}
and, where $\deltav$ assumes the values $\deltav_1$ and $\deltav_2$,
\begin{eqnarray}
S_-(a,\deltav) &=& -a_4^\dagger(\deltav)/\sqrt{18}
+ a_6(\deltav)/\sqrt 6  , \\
S_+(a,\deltav) &=& - a_4(\deltav)/\sqrt{18}
+ a_6^\dagger(\deltav)/\sqrt 6 ,  \\
S_z(a,\deltav) &=& \frac{1}{3} - \frac{a_5^\dagger(\deltav)}{\sqrt {18}}
- \frac{a_5(\deltav)}{\sqrt{18}} - \frac{5n_3(\deltav)}{6} \nonumber \\ &&
- \frac{n_4(\deltav)}{2} - \frac{n_5(\deltav)}{6} + \frac{n_6(\deltav)}{6} \ .
\end{eqnarray}
Thus
\begin{eqnarray}
&& {\cal H} = \sum_\deltav \Biggl[
\frac{k}{36} \biggl( a_4^\dagger(0) a_4(\deltav) \nonumber \\ &&
+ 3 a_6^\dagger(\deltav) a_6(0) +  a_4^\dagger(\deltav) a_4(0)
+ 3a_6^\dagger(0) a_6(\deltav) \biggr) \nonumber \\
&& + \frac{k}{18} \biggl( -5 n_3(\deltav) - 3 n_4 (\deltav)
- n_5 (\deltav) + n_6(\deltav) \nonumber \\ &&
+ n_6(0) - n_5(0) - 3 n_4(0) - 5 n_3(0) \biggr ) 
\nonumber \\ && + \frac{k}{18} \biggl( a_5^\dagger(0) a_5(\deltav)
+ a_5(0) a_5^\dagger (\deltav) \biggr) \Biggr] \ .
\end{eqnarray}

\subsection{c at 0 interacts with b at $\deltav_1$}
%D6

Here
\begin{eqnarray}
S_-(c) &=& - a_4^\dagger(0)/\sqrt {18} + a_6(0)/\sqrt 6 \ , \\
S_+(c) &=& - a_4(0)/\sqrt{18} + a_6^\dagger(0)/\sqrt 6 \ , \\
S_z(c) &=& \frac{1}{3} - \frac{a_5^\dagger(0)}{\sqrt {18}}
- \frac{a_5(0)}{\sqrt{18}} + \frac{n_6(0)}{6} \nonumber \\ &&
- \frac{n_5(0)}{6} - \frac{n_4(0)}{2} - \frac{5n_3(0)}{6}
\end{eqnarray}
and
\begin{eqnarray}
S_-(b,\deltav_1) &=&  2a_4^\dagger(\deltav_1)/\sqrt{18} 
- 2a_6(\deltav_1)/\sqrt 6 \ , \\
S_+(b,\deltav_1) &=& 2a_4(\deltav_1)/\sqrt{18}
- 2a_6^\dagger(\deltav_1)/\sqrt 6\ , \\
S_z(b,\deltav_1) &=& -\frac{1}{6} + \frac{2a_5^\dagger(\deltav_1)}{\sqrt {18}}
+ \frac{2a_5(\deltav_1)}{\sqrt{18}} - \frac{n_3(\deltav_1)}{3}  \nonumber \\ &&
+ \frac{n_5(\deltav_1)}{3} + \frac{2n_6(\deltav_1)}{3} \ .
\end{eqnarray}
Thus
\begin{eqnarray}
&&  {\cal H} = j[ -a_4^\dagger(0) a_4(\deltav) 
- 3 a_6^\dagger({\deltav}) a_6(0) \nonumber \\ && - a_4^\dagger(\deltav) a_4(0)
- 3a_6^\dagger(0) a_6(\deltav) ]/18 \nonumber \\
&& + j [ -4 n_3(\deltav) +4 n_5(\deltav) + 8 n_6(\deltav)
\nonumber \\ && - n_6(0) + n_5(0) + 3 n_4(0) + 5 n_3(0) ]/36 \nonumber \\
&& + j[ -a_5^\dagger(0) a_5(\deltav) -a_5(0) a_5^\dagger(\deltav) ]/9 \ .
\end{eqnarray}

\subsection{Summary}
%D7
Summing all the above contributions we get the Hamiltonian for the
band at energy $J$ for the ferroconfiguration as

\begin{figure} [ht!] 
\vspace*{0.2 in}
\begin{center}
\includegraphics[width=2.6 in]{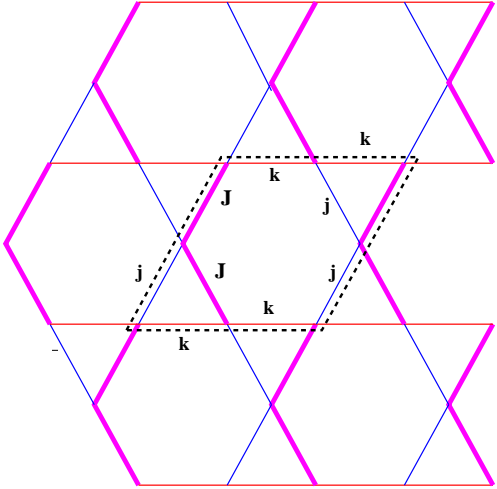} 
\caption{\label{KAG2} As Fig. 1 for a covering with one trimer per unit cell.
In this case the paramagnetic unit cell contains a single trimer. The
magnetic unit cell in the presence of antiferromagnetic trimer
ordering is the same as that in Fig. \ref{KAG} and contains two trimers.}
\end{center}
\end{figure}

\begin{eqnarray}
{\cal H} &=& \sum_{\bf R} \biggl[ 2k[ n_6({\bf R}) -n_5({\bf R})
 -3n_4({\bf R}) -5n_3({\bf R}) \nonumber \\ &&
\hspace*{-0.2 in} -5 n_3({\bf R}_1) -3 n_4({\bf R}_1)
- n_5({\bf R}_1) + n_6({\bf R}_1)]/9 \nonumber \\ && \hspace*{-0.2 in}
+j[ 7n_6({\bf R}) + 5 n_5({\bf R}) + 3 n_4({\bf R}) + n_3({\bf R}) \nonumber \\
&&  \hspace*{-0.2 in}+ n_3({\bf R}_1) + 3 n_4({\bf R}_1) + 5 n_5({\bf R}_1)+7n_6({\bf R}_1)]/18
\nonumber \\ && \hspace*{-0.2 in}
+ \sum_\deltav j[-a_4^\dagger({\bf R}) a_4({\bf R}+ \deltav) -
a_4^\dagger ({\bf R}+ \deltav) a_4({\bf R}) \nonumber \\ && \hspace*{-0.2 in} 
-3 a_6^\dagger({\bf R}) a_6({\bf R}+\deltav)
-3a_6^\dagger({\bf R}+\deltav) a_6({\bf R})  \nonumber \\ && \hspace*{-0.2 in}
-2 a_5^\dagger({\bf R}) a_5({\bf R}+\deltav) 
-2 a_5^\dagger({\bf R}+\deltav) a_5({\bf R}) ]/18 
\nonumber \\ && \hspace*{-0.2 in}
+ k[ a_4^\dagger ({\bf R}) a_4 ({\bf R}+\deltav)
+ a_4^\dagger ({\bf R}+\deltav) a_4 ({\bf R}) \nonumber \\ && \hspace*{-0.2 in}
+ 2a_5^\dagger ({\bf R}) a_5 ({\bf R}+\deltav) 
+ 2a_5^\dagger ({\bf R}+\deltav) a_5({\bf R})/36
\nonumber \\ && \hspace*{-0.2 in}
+ 3a_6^\dagger ({\bf R}) a_6 ({\bf R}+\deltav) 
+ 3a_6^\dagger ({\bf R}+\deltav) a_6({\bf R})]/36 \biggr] \ .
\nonumber \\ &&
\label{EQD7} \end{eqnarray}

\section{Another trimer covering}

In Fig. \ref{KAG2} we show another covering of the Kagom\'e lattice
with trimers.  If the trimers are antiferromagnetically ordered,
then the magnetic unit cell is the same as that of Fig. \ref{KAG} and
one can verify that the spectrum within the ground
manifold is again given by Eqs. (\ref{EQAF}) and (\ref{EQF}).

\end{appendix}


\begin{thebibliography} {99}
\bibitem{MW}
M. Weistein, Phys. Rev. D {\bf 61}, 034505 (2000).
\bibitem{GN}
D. Grohol and D. G. Nocera, Chem. Mater, {\bf 19}, 3061 (2007).
\bibitem{KSB}
M. Kohno, O. A. Starykh, and L. Balents, Nat. Phys. {\bf 3}, 790 (2007).
\bibitem{JAPS}
M. Ishii, H. Tanaka, M. Hori, H. Uekusa, Y. Ohashi, K. Tatani, Y. Narumi,
and K. Kindo, J. Phys. Soc. Jpn, {\bf 69}, 340 (2000).
% Gapped ground state in a Trimer Chain System
\bibitem{FURRER2}
A. Furrer and H. U. G\"udel, J. Magn. Magn. Mater. {\bf 14}, 256 (1979).
\bibitem{FURRER1}
U. Falk, A. Furrer, N. Furrer, H. U. G\'udel, and J. K. Kjems, Phys. Rev.
B {\bf 35}, 4893 (1987).
\bibitem{CB}
Y. Qiu, C. Broholm, S. Ishiwata, M. Azuma, M. Takano, R. Bewley, and
W. J. L. Buyers, Phys. Rev. B {\bf 71}, 214439 (2005).
% Spin Trimer Antiferromagnetism in La4Cu3MnO12
\bibitem{POD}
A. Podlesnyak, V. Pomjakushin, E. Pomjakushina, K. Conder, and A. Furrer,
Phys. Rev. B {\bf 76}, 064420 (2007).
% Magnetic excitations in the spin-trimer compound Ca_3Cu_{3-x}Ni_x(PO4)_4
\bibitem{OKAM} %Diamond chain Cu3Cl6(H2O)2 . 2H8C4SO2
K. Okamoto, T. Tonegawa, Y. Takahashi, and M. Kaburagi, J. Phys.:
Condens. Matter {\bf 11}, 10485 (1999).
\bibitem{HL}
A. Honecker and A. L\"auchli, Phys. Rev. B {\bf 63}, 174407 (2001).
\bibitem{SHL1}
S.-H. Lee, H. Kikuchi, Y. Qiu, B. Lake, Q. Huang, K. Habicht,
and K. Keefer,  Nature Mater. {\bf 6}, 853 (2007).
\bibitem{SHL2}
J.-H. Kim, S. Ji, S.-H. Lee, B. Lake, T. Yildirim, H. Nojiri,
H. Kikuchi, K. Habicht, Y. Qiu, and K. Kiefer, Phys. Rev. Lett.
{\bf 101}, 107201 (2008).
\bibitem{DM1}
I. E. Dzialoshinskii, J. Phys. Chem. Solids {\bf 4}, 241 (1958).
\bibitem{DM2}
T. Moriya, Phys. Rev. {\bf 120}, 91 (1960).
\bibitem{DM3}
T. Yildirim and A. B. Harris, Phys. Rev. B {\bf 73}, 214446 (2006).
\bibitem{DM4}
O. C\'epas, C. M. Fong, P. W. Leung, and C. Lhuillier, Phys. Rev. B
{\bf 78} 140405(R) (2008).
%\bibitem{ITC}
%A. J. C. Wilson, {\it International Tables for Crystallography},
%(Kluwer Academic, Dordrecht, 1995), Vol. A.
\bibitem{ROSE}
M. E. Rose, {\it Elementary Theory of Angular Momentum}
(John Wiley \& Sons, Inc. New York, 1957).
\bibitem{PB}
Note that if one uses the product wave functions of the N\'eel state,
then the phase boundary between ferro- and antiferro-magnetic states
occurs at $k=j$.  For large $J$ the zero-point corrections to
the N\'eel state phase boundary are more severe than those to the
trimer state phase boundary. The effects of such quantum fluctuations
on phase boundaries were studied by
E. Rastelli and A. B. Harris, Phys. Rev. B {\bf 41}, 2449 (1990).
\bibitem{SWAVE}
F. Keffer, in {\it Encyclopedia of Physics: Ferromagnetism}, edited by
S. Fl\"ugge and H. P. J. Wijn (Springer, Berlin, Germany, 1966), p1.
\bibitem{HEINE}
V. Heine, {\it Group Theory in Quantum Mechanics} (Pergamon, New York,
1960) p284.
\bibitem{FJD}
F. J. Dyson, Phys. Rev. {\bf 102}, 1217 (1956); Phys. Rev. {\bf 102},
1230 (1956).
\bibitem{CHAINS}
J. desCloizeaux and J. J. Pearson, Phys. Rev. {\bf 128}, 2131 (1962).
\bibitem{PERT}
T. Oguchi, Phys. Rev. {\bf 117}, 117 (1959)
%\bibitem{HKHH}
%A. B. Harris, D. Kumar, B. I. Halperin, and P. C. Hohenberg, Phys.
%Rev. B {\bf 3}, 961 (1971).
\bibitem{ACTA}
For a survey of the structural properties of polymorphs of Cu$_2$(OH)$_2$Cl
see T. Malcherek and J. Schl\"uter, Acta. Cryst. {\bf B65}, 334 (2009).
\bibitem{WILLS}
A. S. Wills and J.-Y. Henry, J. Phys.: Condens. Matter {\bf 20}, 472206 (2008).

\end{thebibliography}
\end{document}